\documentclass[
 fleqn,
 usenatbib,
]{mnras}

\usepackage{newtxtext,newtxmath}
\DeclareRobustCommand{\VAN}[3]{#2}
\let\VANthebibliography\thebibliography
\def\thebibliography{\DeclareRobustCommand{\VAN}[3]{##3}\VANthebibliography}
\usepackage{graphicx}
\usepackage{dcolumn}
\usepackage{hyperref}
\usepackage{bm}
\usepackage[section]{placeins}
\usepackage{ulem}

\title[DL from 21cm in 3D]{Inferring Astrophysics and Dark Matter Properties from 21cm Tomography using Deep Learning}

\author[S. Neutsch et al.]{
Steffen Neutsch,$^{1}$\thanks{E-mail: steffen.neutsch@hs.uni-hamburg.de}
Caroline Heneka,$^{1}$\thanks{E-mail: caroline.heneka@uni-hamburg.de}
and Marcus Brüggen$^{1}$
\\
$^{1}$University of Hamburg, Hamburger Sternwarte, Gojenbergsweg 112, D-21029 Hamburg, Germany}

\date{\today}
\pubyear{2022}
\begin{document}
\maketitle
\begin{abstract}
21cm tomography opens a window to directly study astrophysics and fundamental physics of early epochs in our Universe's history, the Epoch of Reionisation (EoR) and Cosmic Dawn (CD). Summary statistics such as the power spectrum omit information encoded in this signal due to its highly non-Gaussian nature. Here we adopt a network-based approach for direct inference of CD and EoR astrophysics jointly with fundamental physics from 21cm tomography. We showcase a warm dark matter (WDM) universe, where dark matter density parameter $\Omega_\mathrm{m}$ and WDM mass $m_\mathrm{WDM}$ strongly influence both CD and EoR. 
Reflecting the three-dimensional nature of 21cm light-cones, we present a new, albeit simple, 3D convolutional neural network (\href{https://github.com/stef-neu/3D-21cmPIE-Net}{3D-21cmPIE-Net}) for efficient parameter recovery at moderate training cost. On simulations we observe high-fidelity parameter recovery for CD and EoR astrophysics ($R^2>0.78-0.99$), together with DM density $\Omega_\mathrm{m}$ ($R^2>0.97$) and WDM mass ($R^2>0.61$, significantly better for $m_\mathrm{WDM}<3-4\,$keV). For realistic mock observed light-cones that include noise and foreground levels expected for the Square Kilometre Array, we note that in an optimistic foreground scenario parameter recovery is unaffected, while for moderate, less optimistic foreground levels (occupying the so-called wedge) the recovery of the WDM mass deteriorates, while other parameters remain robust against increased foreground levels at $R^2>0.9$. We further test the robustness of our network-based inference against modelling uncertainties and systematics by transfer learning between bare simulations and mock observations; we find robust recovery of specific X-ray luminosity and ionising efficiency, while DM density and WDM mass come with increased bias and scatter.
\end{abstract}

%%%%%%%%%%%%%%%%%%%%%%%%%%%
\section{Introduction}
\label{sec:introduction}
%%%%%%%%%%%%%%%%%%%%%%%%%%%
\noindent Cosmic Dawn (CD), when the first luminous sources in the Universe form, and the Epoch of Reionisation (EoR), when the first stars and galaxies ionise the surrounding intergalactic medium (IGM), are key eras in our Universe's history. Mapping the spin-flip transition of neutral hydrogen, the so-called 21cm line, can provide us with 3D tomographic light-cones of these first billion years of our Universe. In recent years radio interferometers target the redshifted 21cm radiation to probe the EoR and CD. While current experiments such as the Precision Array for Probing the Epoch of Reionisation (PAPER; \citet{Parsons_2010}), the Murchison Widefield Array (MWA; \citet{Tingay_2013}), the Low Frequency Array (LOFAR; \citet{van_Haarlem_2013}) and the Hydrogen Epoch of Reionisation Array (HERA; \citet{DeBoer_2017}) aim for a statistical detection, the Square Kilometre Array (SKA)\footnote{https://www.skatelescope.org/} is set to provide full 3D tomography using the 21cm line emission.

Besides probing early star and galaxy formation/evolution, 21cm intensity mapping is a promising tool to probe our cosmic concordance model at $z>6$, e.g. the nature of dark energy and/or modifications of gravity~\citep{Heneka:2018kgn,Heneka2018,LiuWP2019,Berti2021} and the nature of dark matter (DM), see for example \citet{Evoli_2014,Sitwell_2014,List2020, FDM2021} and the recent 21cm power spectrum constraints by~\citet{HERA2021}.
The $\Lambda$CDM model, where DM is cold and non-baryonic, agrees well with a wide range of cosmological observations. A DM candidate that is semi-relativistic at creation and is predicted at masses of the order of keV is warm wark matter (WDM)~\citep{WDM1983}. Generally, WDM encompasses thermal relics at few keV that decouple well before ordinary neutrinos and lead to small scale modifications of the cold dark matter power spectrum. A popular possible WDM particle is the sterile neutrino~\citep{Dodelson_1994,Abazajian_2001,PhysRevLett.102.201304,Adhikari_2017}; another possible candidate is the light gravitino, the supersymmetric partner of the graviton, of masses up to few keV~\citep{2005PhRvD..71f3534V}. Depending on its mass, the corresponding free streaming length of thermal WDM suppresses the DM power spectrum and thus structure formation at small scales. 
 This effect has been hypothesised to both aid with the so-called missing satellites problem of too few substructure observed in galaxies~\citep{Moore_1999}, as well as the "too big to fail" problem~\citep{Boylan_Kolchin_2011}. Generally, measurements of the 21cm line during CD and the EoR can provide us with a useful handle on the underlying DM power spectrum and thus (W)DM properties at these redshifts of early structure formation.

So far, summary statistics such as the power spectrum have been extensively used to investigate the impact of WDM on the 21 cm signal (e.g. \citet{Sitwell_2014,10.1093/mnras/stt2431,Carucci_2015,Villanueva_Domingo_2018}). They have been shown to be sensitive to WDM properties. However, the 21cm signal is highly non-Gaussian~\citep{Mellema_2006} and beyond-Gaussian statistics such as the bispectrum (see~\citet{Saxena_2020} for the comparison of bispectra in CDM and WDM scenarios) have been shown to improve on constraints.
Instead of looking 'by-hand' for optimal summary statistics to provide a complete as possible description of the signal, we want to train a deep neural network (NN) for optimal parameter recovery, including the WDM mass, from 21 cm maps. Previously NNs have been applied for study of the 21cm signal, including direct parameter inference from the 21cm power spectrum~\citep{Shimabukuro_2017} and signal emulation~\citep{Kern_2017,Schmit_2017}. More recently, there have been advances for direct inference of 21cm astrophysics and cosmology from imaging and tomography~\citep{Gillet_2019,2020HassanCNN,Mangena_2020,Hort_a_2020,prelogovic2021machine}. So far this approach has not been tested for a WDM universe. 

In this paper we enlarge the parameter space from key astrophysical parameters of CD and the EoR to DM properties and test the application of deep learning for joint parameter recovery of astrophysical and WDM parameters from 21cm light-cones. Besides searching for an optimal and at the same time simple (meaning fast converging during training and well-interpretable) network model for the joint inference task, we are interested in how well an NN can infer WDM properties, given it is presented with the full non-Gaussian information encoded in 21cm light-cones.

The paper is organised as follows: We start with a brief overview of the creation of our database of 21cm light-cones in Section~\ref{sec:simulations}. We then present our findings from our search for an optimal network architecture in Section~\ref{sec:architecture}. In Section \ref{sec:results} we present our results for parameter inference for different network architectures, for different levels of foregrounds, and for both an astrophysics plus WDM and astrophysics-only set of parameters. Moreover, we test the robustness of our results by transfer learning between simulations and mock observed signals. Finally, in Section~\ref{sec:interpretation} we take a look at saliency maps and filter representations to understand how our best-performing 3D CNN operates.

%%%%%%%%%%%%%%%%%%%%%%%%%%%
\section{Database of 21 cm light-cones}
\label{sec:simulations}
%%%%%%%%%%%%%%%%%%%%%%%%%%%
In this section we describe the creation of the database of 3D light-cones of 21cm brightness temperature fluctuations $\delta T_\mathrm{b}\left(\bf{x},\nu\right)$, with on-sky coordinates, $\bf{x}$, and frequency, $\nu$, which we use for training our neural networks. For simulation of the light-cones we use the publicly available semi-numerical code 21cmFAST v3\footnote{https://github.com/21cmFAST/21cmFAST}~\citep{Mes07,Mesinger2010,Murray2020}. The light-cones are generated from evolving coeval cubes with redshift as follows. 21cmFAST generates density and velocity initial conditions in Lagrangian space. The density and velocity fields are then evolved at first and second order perturbation theory using the Zel'dovich approximation~\citep{1970Zel}. 
Collapsed regions are identified within an excursion-set procedure by applying a top-hat filter at decreasing scales. As described in~\citet{2004reionexcursion} a region is fully ionised when the fraction of collapsed matter contributing to star formation in the region $f_\mathrm{coll}\left( \boldsymbol{x} ,R,z \right)$ exceeds the inverse ionisation efficiency $\zeta^{-1}$. Partially ionised regions are included with an ionisation fraction of $\zeta f_\mathrm{coll}\left( \boldsymbol{x} ,R,z \right)$~\citep{Mes07}. Given the 21 cm signal during reionisation generally depends on the gas spin temperature $T_{\rm S}$ if we drop the simplifying assumption that the IGM was already pre-heated before reionisation~\citep{Greig2017,2020MNRAS.496..581H}, we do not assume the so-called post-heating regime of $\bar{T}_{\rm S} \gg T_{\rm CMB}$ but instead evolve the spin temperature field $T_\mathrm{S}$ for each redshift.

%%%%%%%%%%%%%%%%%%%%%%%%%%%
\subsection{Simulation and Choice of Parameters}\label{sec:parameters}
%%%%%%%%%%%%%%%%%%%%%%%%%%%

\noindent Our goal is to infer WDM properties from 21cm light-cones, jointly with a standard-set of astrophysical parameters that govern the expected 21cm signal. For this proof of concept study we train our networks on a total of six different parameters. This includes four key astrophysical - and two DM parameters. 
For the astrophysics-only scenario (see Section~\ref{sec:datasets}) quantities were derived in co-moving units
assuming flat $\Lambda$CDM: ($\Omega_\mathrm{m}$, $\Omega_\mathrm{b}$, $n_\mathrm{s}$, $H_0$, $\sigma_8$) = (0.31, 0.049, 0.97, 68 km s$^{-1}$ Mpc$^{-1}$, 0.81), in accordance with \textit{Planck}~\citep{Planck2018} results; cosmological parameters were held fixed to fiducial values with the exception of $\Omega_\mathrm{m}$ and $m_\mathrm{WDM}$ for our WDM scenario.

For light-cone generation and training we investigate the following ranges for the two DM parameters:
\begin{itemize}
    \item Warm dark matter mass $m_\mathrm{WDM}\epsilon [0.3,10]\,$keV. 
    \newline Current constraints e.g. using the Lyman-alpha forest include $m_\mathrm{WDM}>3.5\,$keV (95$\%$ CL) from combining results from HIRES, MIKE and X-shooter spectrographs~\citep{Ir_i__2017} and $m_\mathrm{WDM}>5.3\,$keV (95$\%$ CL) using BOSS and eBOSS data~\citep{Palanque_Delabrouille_2020}, as well as for a more conservative treatment of systematic uncertainties $m_\mathrm{WDM}>1.9\,$keV~\citep{10.1093/mnras/stab192}. To allow for a wide range of parameter behaviour we use a relatively small lower limit of $m_\mathrm{WDM}=0.3\,$keV. We also test comparably high WDM masses that make the model look increasingly similar to CDM due to the decreased free-streaming scale of WDM.
    \item Dark matter density parameter $\Omega_\mathrm{m}\epsilon [0.2,0.4]$. 
    \newline It crucially controls structure formation, and we choose a fairly large allowed range encompassing current constraints such as from the \textit{Planck} satellite~\citep{Planck2018}.
\end{itemize}

The chosen ranges for four remaining key astrophysical parameters are as follows:
\begin{itemize}
    \item The minimum virial temperature $T_\mathrm{vir}\epsilon [10^{4},10^{5.3}]\,$K. It determines the minimum virial temperature of halos for cooling to be efficient for star formation. The lower limit corresponds to a minimum temperature for efficient atomic cooling, the upper limit is motivated by observations of high redshift Lyman break galaxies~\citep{Greig2015}.
    \item The ionisation efficiency $\zeta\epsilon [10,250]$. $\zeta$ is a composite parameter given by $\zeta=30\left(\frac{f_\mathrm{esc}}{0.3}\right)\left(\frac{f_\mathrm{*}}{0.05}\right)\left(\frac{N_\mathrm{\gamma/b}}{4000}\right)\left(\frac{2}{1+n_\mathrm{rec}}\right)$, with the fraction of ionising photons escaping into the IGM $f_\mathrm{esc}$, the fraction of galactic gas in stars $f_\mathrm{*}$, the number of ionising photons per baryon in stars $N_\mathrm{\gamma}$ and the typical number density of recombinations for hydrogen in the IGM $n_\mathrm{rec}$. 
    The chosen parameter range for $\zeta$ allows for a wide range of allowed reionisation scenarios.
    \item The specific X-ray luminosity (with energies < 2$\,$keV) per unit star formation rate that escapes host galaxies $L_\mathrm{X}\epsilon [10^{38},10^{42}]$erg s$^{-1}$ M$^{-1}_\mathrm{\odot}$ yr. 
    The parameter range encompasses limits from observations of star-forming galaxies and high-redshift simulations~\citep{Greig2017}.
    \item The X-ray energy threshold for self absorption by host galaxies $E_0\epsilon [100,1500]\,$ eV; X-rays of energies below $E_0$ do not escape the host galaxy. The parameter range is motivated by the column density of the ISM in simulated high-redshift galaxies~\citep{Das2017}.
\end{itemize}

Note that for network training and evaluation in the following all parameter labels are normalised to [0,1] for improved model performance to avoid parameters to be weighted differently according to their magnitude.

We produce light-cones by randomly sampling the given parameter ranges. We use a box size of 200 Mpc with a resolution of 1.43 Mpc. For the third redshift dimension we simulate a range of $z=5-35$.
Note that $\Omega_\mathrm{m}$ controls the relative importance between DM and DE for flat cosmologies. Consequently a change in $\Omega_\mathrm{m}$ leads to a change in length of the produced light-cones both in terms of Mpc and number of pixels. To standardise the input size for our network, the first 2350 pixels of each light-cone are kept, meaning that a light-cone of $\Omega_\mathrm{m}=0.4$ reaches $z=35$, while light-cones with lower $\Omega_\mathrm{m}$ end at lower redshifts.
Furthermore, when producing light-cones by randomly sampling over the given parameter ranges, sometimes light-cones with e.g. unrealistically late or early reionisation in comparison to current constraints are generated. We therefore restrict the light-cones trained on in terms of the optical depth and low-redshift neutral fraction. We require $\tau$ to be within $5\sigma$ of the measurement of $0.054\pm0.007$ from the \textit{Planck} satellite~\citep{Planck2018}, as well as the IGM to be mostly ionised by redshift 5 in order to not contradict Ly$\alpha$ forest observations. Still wanting to include a wide range of potential models we only require the IGM mean neutral fraction at redshift 5 to be below 0.1.

%%%%%%%%%%%%%%%%%%%%%%%%%%%
\subsection{Creation of Mock Observed Light-Cones}\label{sec:mocks}
%%%%%%%%%%%%%%%%%%%%%%%%%%%
We create mock versions of our simulated light-cones with the publicly available code 21cmSense\footnote{https://github.com/jpober/21cmSense}~\citep{Pober_2013,Pober_2014}. We directly transform the simulated light-cones from the previous section into SKA mock observed light-cones. 
To do so, we take the light-cone array of coevally evolved simulation boxes at fixed redshifts that 21cmFAST creates. To create mock light-cones the light-cone is again split into coeval boxes at certain redshift values. Two connected boxes are split at the midpoint corresponding to their respective redshifts. 
For each box (or frequency slice) the thermal noise is then calculated using 21cmSense and the resulting noise is added to the Fourier-transformed box by randomly drawing from a zero-mean Gaussian with variance as derived for thermal noise. The result is then transformed back to real space and the new mock light-cone rebuilt.

For thermal noise we assume 1080 h of integrated SKA-Low stage 1 observations, for tracked scans of 6h each, with instrument characteristics and baseline distribution as described in the SKA1 System Baseline Design document.\footnote{https://astronomers.skatelescope.org/wp-content/uploads/2016/05/SKA-
TEL-SKO-0000002$\_$03$\_$SKA1SystemBaselineDesignV2.pdf} 21cmSense allows for three foreground settings: optimistic ('opt'), moderate ('mod') and pessimistic. In the optimistic foreground scenario the 21cm foreground wedge in k-space only covers the primary field-of-view of the instrument. In the moderate foreground scenario it extends $0.1\,$Mpc$^{-1}$ beyond the horizon limit. In the pessimistic foreground scenario it further extends beyond the moderate foreground scenario due to incoherently added baselines. Here in our analysis we focus on noise levels that assume the optimistic 'opt' and moderate 'mod' foreground scenarios. Examples for opt and mod mock light-cones compared to a bare simulation are shown in Figure~\ref{fig:3D_Architecture}. While opt mock light-cones mainly exhibit noise contamination at higher redshifts, mod mock light-cones also exhibit noise contamination towards lower redshifts and loose information about the absolute brightness temperature (similarly to the scenario of absolute values lost in interferometric measurements and the use of mean averaged light-cones).

%%%%%%%%%%%%%%%%%%%%%%%%%%%
\subsection{2D and 3D Data for Training, Validation, and Test}\label{sec:datasets}
%%%%%%%%%%%%%%%%%%%%%%%%%%%
\noindent We produce a database of 5050 light-cones in agreement with parameter ranges and constraints as detailed in Section~\ref{sec:parameters} and which varies our chosen set of two DM and four key astrophysical parameters. For training of the networks we split the dataset into 3600 light-cones for training, 450 for validation and 1000 for the test set. The size of our test set matches previous works (e.g. \citet{Gillet_2019,zhao2021simulationbased,prelogovic2021machine}), allowing for direct comparison. For our 2D CNN and ResNet we extract one 2D slice per light-cone  by slicing along redshift and one spatial dimension. The resulting slices have a size of $140\times 2350$ pixels. As the LSTM architecture requires a series of 2D slices, our optimal setup takes a series of slices in redshift steps of 0.5 (for $\Omega_\mathrm{m}=0.3$ and starting from z=5). We produce a second and third database of opt mock and mod mock light-cones from this database of bare simulated light-cones.
Analogous to the databases created for our full set of parameters, we produce a fourth database containing 5050 light-cones to exclusively train our best-performing 3D CNN on the astrophysical parameters $L_\mathrm{X}$, $E_0$, $T_\mathrm{vir}$ and $\zeta$ in a CDM universe. We thus can estimate how the performance of our 3D CNN is affected by the addition of DM parameters. For compatibility with previous work the $\tau$ and IGM mean neutral fraction constraints are not applied here. We later refer to this database as our 'astro-only' dataset.

%%%%%%%%%%%%%%%%%%%%%%%%%%%
\section{The optimal Neural Network Architecture for 3D 21cm tomography}
\label{sec:architecture}
%%%%%%%%%%%%%%%%%%%%%%%%%%%

Our goal is to estimate directly a joint parameter set of astrophysical and DM parameters from 3D 21cm light-cones, consisting of 2D imaging in the sky-plane and an additional frequency (redshift) dimension. For training and validation of our networks we generated a database of 21cm light-cones as described in the previous section. To identify the optimal architecture for our parameter inference, we investigated a range of network variants whose underlying architectural concepts match 3D tomographic data (2D spatial slices and a third frequency dimension) as an input.

We start by stressing that our best-performing network is a full 3D, but relatively simple, Convolutional Neural Network (CNN) that takes as input full 3D light-cones. Two further candidate networks, a standard 2D CNN and a Residual Neural Network (ResNet), take 2D slices of one spatial and the frequency dimension to perform 2D convolutions on such sliced light-cones. Another candidate network exploits the idea of two correlated spatial dimensions and and a frequency dimension in form of a time-, or redshift-, sequence. Input is a series of 2D slices that are connected by long and short-range connections in form of a Long Short Term Memory (LSTM) network.

In the following, we will give a brief overview of our different candidate architectures for the joint estimate of astrophysical and DM parameters from 3D light-cone data. All architectures have been optimised with a thorough grid search of possible hyperparameter choices with regards to number of layers and filters, filter kernel sizes, dense layer nodes, activation functions, regularisation techniques (dropout, pooling, batch normalisation), batch size, optimizer, loss function and learning rate. We note that we find the Adam optimizer~\citep{kingma2017adam} with AMSgrad~\citep{AMSGrad} and the mean squared error loss function to be the optimal choice for all of our architectures. Additionally, we use a few common techniques to improve convergence. We use a learning rate reduction of a factor of 0.5 after a plateau of 5 epochs without the validation loss improving. After a plateau of 10 epochs we stop the training and restore the weights from the epoch with the lowest validation loss before applying the neural network to the test set. Finally, all parameter labels (two DM and four astrophysical parameters, see section~\ref{sec:parameters}) are normalised to a range of [0,1] to ensure equal training on the different parameters and improve performance of our network models.

We start by presenting our conclusions from testing and optimising our three 2D network architectures, the 2D CNN, the ResNet and the LSTM, and finish by presenting our 3D network optimised for 21cm light-cones. The best-performing 3D architecture is shown in Figure~\ref{fig:3D_Architecture}, for the other candidate architectures we refer to appendix \ref{sec:2DNN Architectures}. All networks are implemented, trained and evaluated using the Python API Keras,\footnote{https://keras.io} based on TensorFlow.\footnote{https://github.com/tensorflow/tensorflow}

%%%%%%%%%%%%%%%%%%%%%%%%%%%
\subsection{A 2D CNN for 21cm imaging}
%%%%%%%%%%%%%%%%%%%%%%%%%%%
CNNs are one of the most commonly used architectures for image interpretation tasks. They have already been used for deriving properties from 21cm images of reionisation (see for example \citet{Gillet_2019}, \citet{Mangena_2020}, \citet{zhao2021simulationbased}). The idea is to use convolutional layers to read out local features. Dense layers then regress output parameters of interest, in our case our set of astrophysical and DM parameters. In addition to standard convolutional layers we require additional measures against overfitting. Overfitting usually manifests itself in decreased accuracy on validation and test datasets as compared to the training set. Typical measures are dropout, max pooling, average pooling and batch normalisation (\cite{ioffe2015batch}). For our 2D CNN analysis of light-cone slices we find these to significantly increase our performance. A global average pooling (GAP) layer after the final convolutional layer performed best. The GAP layer calculates the average value over each filter and only passes that on to the next layer. While information loss with dropout or batch normalisation has been too large, GAP provided a good balance between information lost by averaging and avoidance of overfitting when using a fairly large number of 128 filters before applying GAP. For our 2D CNN a quite simple architecture with 3x3 filter kernels and the application of Rectified Linear Units~\citep[ReLu,][]{10.5555/3104322.3104425}) activation to all hidden layers yields best results. Exception is the final dense layer without activation function. Not normalising our input image performs best. We used a learning rate of $10^{-4}$ and train the 2D CNN for 100 epochs. Training on a single NVIDIA K80 GPU requires 0.05 hours per epoch. The 2D CNN architecture is summarised in Table~\ref{tab:2DCNN_architecture}.

%%%%%%%%%%%%%%%%%%%%%%%%%%%
\subsection{A ResNet for 21cm imaging}
%%%%%%%%%%%%%%%%%%%%%%%%%%%
To improve performance of a CNN, one way is to simply add more convolutional layers. This however has a major flaw: The network training becomes more difficult, with e.g. gradients for the updating of weights becoming smaller the more layers we use, potentially causing a vanishing gradient problem. One way to solve this is by using skip connections in a so-called ResNet~\citep{He2016DeepRecognition}. A skip connection directly connects non-adjacent layers by adding their output before passing it on. With this in mind we increase the number of convolutional layers as compared to our 2D CNN by a factor of two. For our ResNet we introduce two new features: The first one doubles the kernel size of the first layer, as this helped performance. The second feature is the use of batch normalisation. We suspect that the resulting smoothing of the loss curve makes it more likely to find a sufficiently 'good' local minimum. Analogous to our 2D CNN we use 3x3 kernels for all convolutional layers apart from the first and we refrain from normalising our input values. From the second layer on we group our network in blocks of two layers. The input of each block is connected to its output with a skip connection. Compared to the 2D CNN we use a higher learning rate of $6\times 10^{-4}$. For the results presented our ResNet stopped training after 65 epochs, with the validation loss not improving for 10 epochs. As a result of the deeper architecture, training the ResNet on a single NVIDIA K80 GPU takes about 0.18 hours per epoch, which is considerably longer than the 2D CNN. Our ResNet architecture is presented in Table~\ref{tab:ResNet_architecture}.

%%%%%%%%%%%%%%%%%%%%%%%%%%%
\subsection{An LSTM to capture time-dependence}
%%%%%%%%%%%%%%%%%%%%%%%%%%%
LSTM networks are tailored for sequences of data, such as videos, or texts where time evolution of the signal is relevant for interpretation. Their core feature is the LSTM cell (\citet{HochSchm97}). It stores information from each time step to incorporate time- or sequence-dependent features in the learning. With this in mind we build our LSTM architecture on the idea of exploiting directly the time-dependency of the light-cones. Our LSTM is not bidirectional; it also follows the forward time direction. We create a sequence of 39 spatial 2D slices from each light-cone. A small 2D CNN is applied to each element of the sequence to convolve the spatial features. The results of each time step are connected by an LSTM cell, thus reflecting the time-dependence of the 21cm light-cones. Our optimised LSTM architecture is build with 2D CNNs applied to each time step consisting of a 5x5 convolution, a max pooling layer, a 3x3 convolution and GAP. We use the Swish activation function (\citet{Swish}), apart from the final dense layer which has no activation function, and min-max normalise the input. We start with a learning rate of $4\times 10^{-4}$ and train for 100 epochs, with a training time of 0.10 hours per epoch when training on a single NVIDIA K80 GPU. The best-performing LSTM architecture is summarised in Table~\ref{tab:LSTM_architecture}.

%%%%%%%%%%%%%%%%%%%%%%%%%%%
\subsection{A 3D CNN for full 21cm tomography}\label{sec:3DCNN}
%%%%%%%%%%%%%%%%%%%%%%%%%%%

%%%%%%%%%%%%%%%%%%%%%%%%%%%
\begin{figure*}
    \centering
    \includegraphics[width=\textwidth]{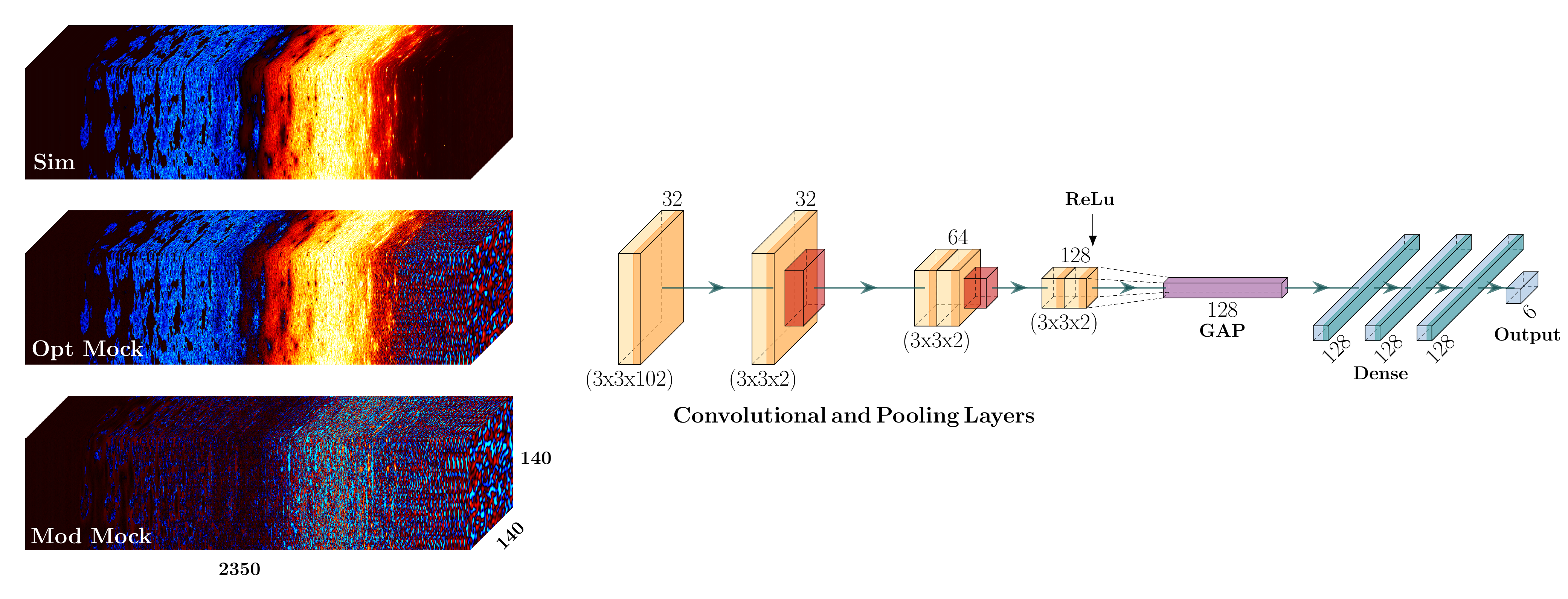}
    \caption{Schematic representation of our best-performing 3D CNN architecture (using code from \citet{PlotNeuralNet}). The network takes as input a 21cm light-cone (left) and outputs a set of inferred parameters. The orange boxes correspond to layers of the network; note that they are four dimensional and therefore the filter dimension is not shown. Instead the number of filters is displayed above each convolutional layer. The blue boxes depict the fully-connected dense part of the network. The darker area of the orange and blue boxes symbolizes the ReLu activation function.
    }
    \label{fig:3D_Architecture}
\end{figure*}
%%%%%%%%%%%%%%%%%%%%%%%%%%%
\noindent While achieving satisfactory results for our 2D networks, a 3D architecture yielded the best performance. Our 3D CNN displays a relatively simple architecture and small number of parameters, while converging after few epochs and a short per epoch training time. A GAP layer against overfitting secures fast convergence. Importantly, the filter kernel of the first convolutional layer is tailored to capture comparably smaller-scale fluctuations spatially and larger-range modulations in frequency, or time (more representative of the global 21cm signal), which significantly speeds up training and improves regression results.

%%%%%%%%%%%%%%%%%%%%%%%%%%%
\subsubsection{Details on 3D architecture, and its optimisation}\label{sec:3DCNNbuild}
%%%%%%%%%%%%%%%%%%%%%%%%%%%
A schematic representation of our 3D CNN is shown in Figure~\ref{fig:3D_Architecture}; Table~\ref{tab:3DCNN_architecture} summarises the architecture. As mentioned, we tailor the 3D CNN architecture to the time-dependency of the 21cm light-cones that follow the evolution of the global 21cm signal. For this purpose filters with a kernel size of $(3,3,102)$ are applied in the first convolutional layer. This enables the network to learn filters specifically designed to detect spatial fluctuations evolving in the redshift direction. We take a more detailed look at the structure of some exemplary filters in Section~\ref{sec:Filter}. Moreover, a stride of $(1,1,102)$ reduces the dimensionality for the following layers, and thus reduces required computational resources and training time. Similar to the 2D networks we use a GAP layer to alleviate overfitting. Compared to other measures against overfitting, this enables the network to converge after fewer epochs while performing better. Some small adjustments to our hyperparameter choices that we describe in the following further optimise the network. We min-max normalise the input with the maximum and minimum pixel values in the full light-cone dataset. The filter size of the hidden convolutional layers is set to (3x3x2) and our max pooling layers are only applied to the two spatial dimensions. The ReLu activation function is used for hidden layers and a sigmoid activation for the final dense layer. This has the side effect of forcing all predictions to lie within the parameter range trained on. 

%%%%%%%%%%%%%%%%%%%%%%%%%%%
\subsubsection{Training of our 3D network}\label{sec:3DCNNtrain}
%%%%%%%%%%%%%%%%%%%%%%%%%%%
The 3D CNN has been trained for 35 epochs on our database of simulated light-cones for bare simulation data, and 30 epochs for mock observed light-cones. Training on mock observed light-cones is slightly slower than training on the bare simulations. This is likely caused by the large number of zero value pixels in most simulation light-cones. For training we use a learning rate of $4\times 10^{-4}$ and set the batch size to 8. The batch size controls the number of light-cones which the neural network reads in before performing a weight update. A smaller batch size results in more frequent weight updates and therefore may lead to larger fluctuations. A larger batch size can lead to a higher degree of generalisation, meaning the neural network may loose some of its ability to interpolate. Using a lower batch size led to a slightly worsened performance. Finally, we note that the training on one NVIDIA K80 GPU takes about 0.33 hours per epoch.

%%%%%%%%%%%%%%%%%%%%%%%%%%%
\begin{table}
    \centering
    \begin{tabular}{ll}
         Layer&Shape\\ \hline
         Input Layer &(140,140,2350,1)
         \\3x3x102 Conv3D &(138,138,23,32)
         \\3x3x2 Conv3D &(136,136,22,32)
         \\2x2x1 Max Pooling &(68,68,22,32)
         \\3x3x2 Conv3D &(66,66,21,64)
         \\1x1x0 Zero Padding &(68,68,21,64)
         \\3x3x2 Conv3D &(66,66,20,64)
         \\2x2x1 Max Pooling &(33,33,20,64)
         \\3x3x2 Conv3D &(31,31,19,128)
         \\1x1x0 Zero Padding &(33,33,19,128)
         \\3x3x2 Conv3D &(31,31,18,128)
         \\Global Average Pooling &(128)
         \\Dense &(128)
         \\Dense &(128)
         \\Dense &(128)
         \\Dense &(6)\\ \hline
         Number of Parameters: 651,526
    \end{tabular}
    \caption{3D CNN Model Summary; see Figure~\ref{fig:3D_Architecture} for a schematic overview.}
    \label{tab:3DCNN_architecture}
\end{table}
%%%%%%%%%%%%%%%%%%%%%%%%%%%

%%%%%%%%%%%%%%%%%%%%%%%%%%%
\section{Results}
\label{sec:results}
%%%%%%%%%%%%%%%%%%%%%%%%%%%
In this section we showcase the ability of our trained neural networks to infer key astrophysical and DM parameters, with emphasis on our best-performing 3D 21 cm parameter inference network (3D-21cmPIE-Net).

To assess performance we calculate the coefficient of determination $R^{2}$ for each parameter on all light-cones of the test set evaluated. The coefficient of determination is defined as
\begin{equation}\label{eq:R2}
R^{2}=1-\frac{\Sigma^\mathrm{n}_\mathrm{i=1}(y_\mathrm{i,true}-y_\mathrm{i,pred})^{2}}{\Sigma^\mathrm{n}_\mathrm{i=1}(y_\mathrm{i,true}-\bar{y})^{2}} . 
\end{equation}
Here $y_\mathrm{i,true}$ is the true value or label of a parameter for light-cone i, $y_\mathrm{i,pred}$ is the prediction for the same parameter for light-cone i and $\bar{y}$ is the average of all labels for this parameter over all light-cones of the test set. All parameter labels are normalised to [0,1]. If the network predicts the value $\bar{y}$ of a parameter for each light-cone, then it is not gathering any information about that parameter. In this case the $R^{2}$ value is zero. The $R^{2}$ value for said parameter will be one, if exactly the true value of a parameter is predicted for each light-cone.

We start with a brief overview of our results for astrophysical and DM parameter inference with 2D architectures trained on simulated light-cones (\ref{sec:2Dresults}), before focusing on the best-performing 3D CNN architecture. We present our findings for inference with the 3D CNN for bare simulated (\ref{sec:3D_Results}), mock observed light-cones with optimal and moderate foreground assumptions (\ref{sec:MockResults}) as well as compare to the inference of astrophysical parameters only (\ref{sec:AstroResults}). We finish with a demonstration of transfer learning where we apply the 3D CNN trained on bare simulated light-cones on mock observed light-cones, and vice-versa, in order to evaluate the robustness of our model for example against inaccuracies in noise and foreground models assumed.

%%%%%%%%%%%%%%%%%%%%%%%%%%%
\renewcommand{\arraystretch}{1.7}
\setlength{\tabcolsep}{17pt}
\begin{table*}
    \centering
    \begin{tabular}{l|cccccc|c}
         Network Model &$m_\mathrm{WDM}$&$\Omega_\mathrm{m}$&$L_\mathrm{X}$&$E_0$&$T_\mathrm{vir}$&$\zeta$&loss\\ \hline
         A. 2D CNN&0.499&0.972&0.980&0.705&0.767&0.957&0.0146
         \\B. ResNet&0.454&0.953&0.979&0.668&0.730&0.935&0.0167
         \\C. LSTM&0.503&0.978&0.984&0.822&0.771&0.964&0.0129
         \\D. 3D CNN$_\mathrm{Sim}$& 0.613 &0.974& \bf{0.986} & \bf{0.826} &0.776& \bf{0.975} & \bf{0.0110}
         \\E. 3D CNN$_\mathrm{OptMock}$&\bf{0.626}& \bf{0.981} &0.980&0.795& \bf{0.795} &0.973&0.0111
         \\F. 3D CNN$_\mathrm{ModMock}$&0.414&0.978&0.971&0.691&0.673&0.913&0.0176
    \end{tabular}
    \caption{Overview of coefficient of determination $R^{2}$ for astrophysical and DM parameters, being warm dark matter mass $m_\mathrm{WDM}$, dark matter density $\Omega_\mathrm{m}$, specific X-ray luminosity $L_\mathrm{X}$, X-ray energy threshold $E_0$, virial temperature $T_\mathrm{vir}$ and ionisation efficiency $\zeta$, as well as test loss values for a range A. to F. of network models. The models represent A. the 2D CNN trained on simulations, B. the 2D ResNet trained on simulations, C. the LSTM trained on simulations, D. the 3D CNN trained on simulations, E. the 3D CNN trained on mock light-cones with optimistic foreground settings, and F. the 3D CNN trained on mock light-cones with moderate foreground settings. In bold we highlight the best-performing model for each column (parameter). Note that generally the 3D CNN on bare simulated light-cones and on mock light-cones with optimal foreground treatment perform best.}
    \label{tab:Results}
\end{table*}
%%%%%%%%%%%%%%%%%%%%%%%%%%%

%%%%%%%%%%%%%%%%%%%%%%%%%%%
\subsection{Parameter Inference from 21cm simulated light-cones: Astrophysics and WDM Properties}
\label{sec:SimulationResults}
%%%%%%%%%%%%%%%%%%%%%%%%%%%

%%%%%%%%%%%%%%%%%%%%%%%%%%%
\subsubsection{Results for 2D Imaging Networks}\label{sec:2Dresults}
%%%%%%%%%%%%%%%%%%%%%%%%%%%

The $R^2$ values of the test set prediction of astrophysical and DM parameters inferred with our 2D CNN, ResNet and LSTM architectures are shown in row A to C in Table~\ref{tab:Results}. For direct comparison of the overall performance the table includes as well the minimum mean squared loss value achieved for the test set by each network. A lower loss value means that the model has a better average prediction across all parameters. The LSTM generally performs best, especially when estimating the X-ray energy threshold $E_0$, and similar to the other networks for the remaining parameters. Generally, across 2D networks, for the matter density $\Omega_\mathrm{m}$, X-ray luminosity $L_\mathrm{X}$ and ionising efficiency $\zeta$ very good inference results (high $R^2 > 0.9$) are achieved, while the warm dark matter mass $m_\mathrm{WDM}$ turns out to be more challenging at $R^2\sim 0.5$. We note though that inference results for $m_\mathrm{WDM}$ are significantly WDM mass-dependent, as is the case for 3D inference described in the following section. We presume that the LSTM performs best since it accounts for the time dependence of the 21cm maps. These results also indicate that increasing the depth of the architecture with a ResNet does not improve performance, further motivating our findings for a relatively small and simple 3D CNN architecture.

%%%%%%%%%%%%%%%%%%%%%%%%%%%
\subsubsection{Results for 3D Tomography}
\label{sec:3D_Results}
\begin{figure*}
\includegraphics[width=0.411\textwidth]{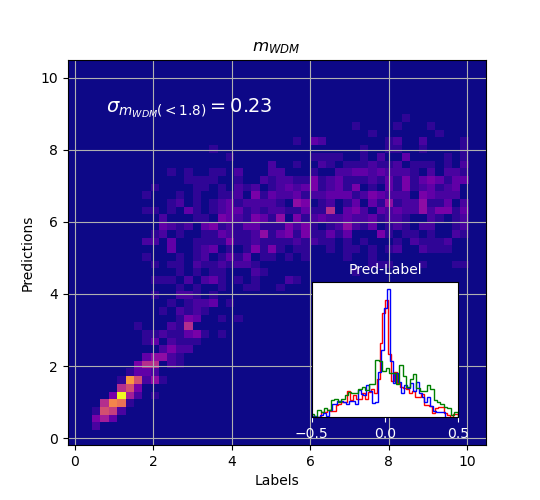}
\includegraphics[width=0.506\textwidth]{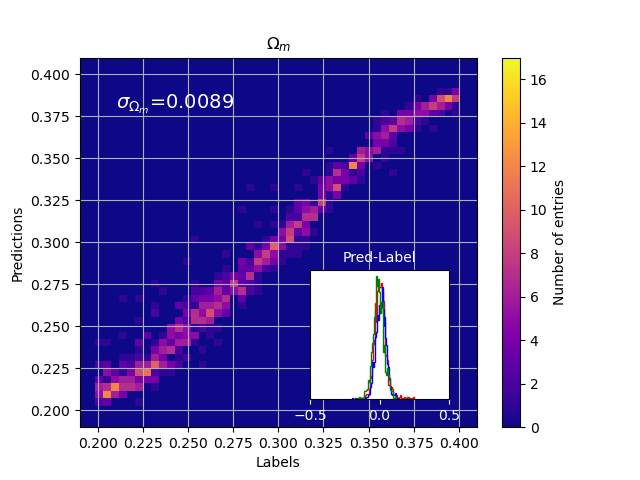}
\includegraphics[width=0.411\textwidth]{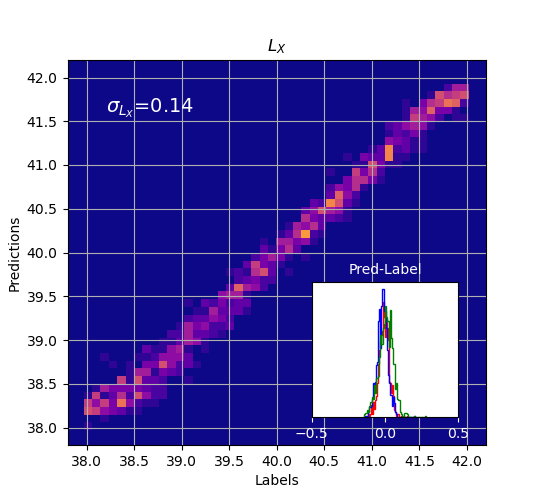}
\includegraphics[width=0.506\textwidth]{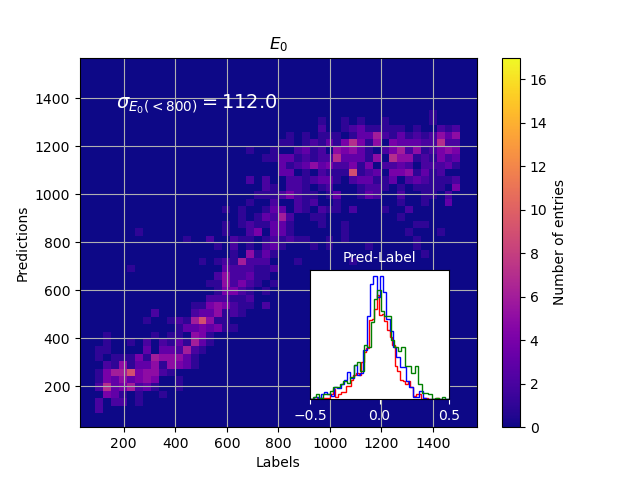}
\includegraphics[width=0.411\textwidth]{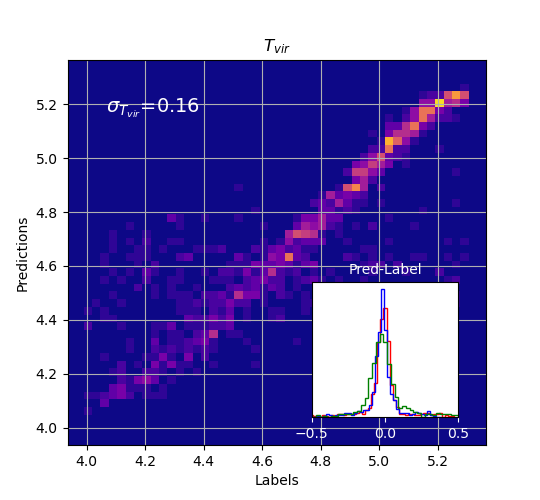}
\includegraphics[width=0.506\textwidth]{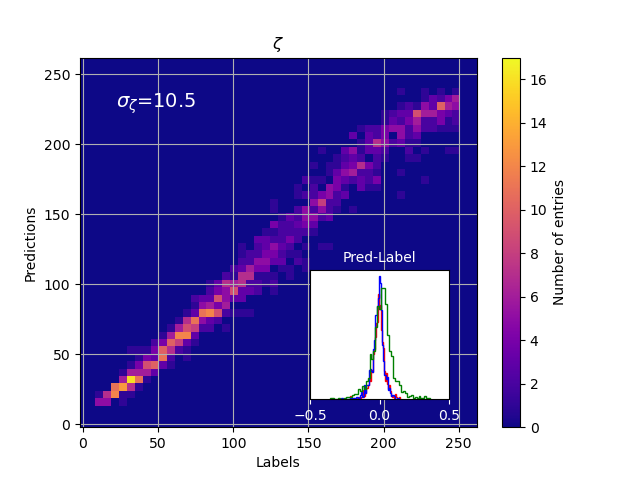}
\caption{Scatter plots of predicted against true parameter labels for each parameter as indicated, evaluated on the test dataset, to showcase the accuracy of the neural network trained on simulated light-cones (large panels). The inlets show the histograms of differences between the predictions and true labels for simulated 21cm light-cones (red), opt mock (blue) and mod mock light-cones (green) for each respective parameter; for details on the mock generation see Section~\ref{sec:mocks}.}
\label{fig:Results}
\end{figure*}

%%%%%%%%%%%%%%%%%%%%%%%%%%%
Here we discuss in more detail findings from our 3D CNN, that shows the best performance as compared to the range of 2D architectures investigated. 
We note the loss curves of our 3D network indicate it to converge well without overfitting. For example when trained on simulated light-cones the network reaches a loss of 0.0094 for the training set, 0.0118 for the validation set and 0.0110 for the test set; see Section~\ref{sec:3DCNNtrain} for more information on the training of our 3D network.
\\ The $R^2$ values of the test set predictions by our 3D CNN are displayed in row D in Table~\ref{tab:Results}. The best results for each parameter across all architectures and datasets are highlighted in bold. Our 3D CNN outperforms our 2D architectures on the simulation dataset, while the LSTM is our best-performing 2D network. We find the largest difference between the LSTM and our 3D network for the parameter $m_\mathrm{WDM}$, with the 3D CNN providing more accurate predictions for low-mass WDM. Again, as for 2D architectures in the previous section, the matter density $\Omega_\mathrm{m}$, X-ray luminosity $L_\mathrm{X}$ and ionising efficiency $\zeta$ are inferred extremely well (high $R^2 > 0.97$), while the warm dark matter mass $m_\mathrm{WDM}$ remains the most challenging parameter. For the virial temperature $T_\mathrm{vir}$ and the X-ray energy threshold $E_0$ we reach $R^2$ values of >0.77 and >0.82, with higher scatter at high and low parameter values, respectively. The scatter plots for the test set predictions of the 3D CNN as compared to the true values are shown in Figure \ref{fig:Results}. For most parameters the scatter plots show some kind of an s-shape and slightly worse results for very high or low values. This has been seen in previous work and can be attributed to the sharp boundaries of the prior parameter ranges~\citep{prelogovic2021machine}. 
Note that we have conservatively chosen wide prior parameter ranges and thus should be safe from the bias caused by this effect for realistic models.

Our results for the parameters $L_\mathrm{X}$, $E_0$ and $\zeta$ are comparable to previous studies where astrophysical parameters alone where inferred~\citep{Gillet_2019,prelogovic2021machine}; see also the discussion in Section~\ref{sec:AstroResults}.  We therefore focus here our discussion on the remaining three parameters.
When training on an astro-only dataset our 3D CNN achieves $R^2$ values above 0.99 for $T_\mathrm{vir}$ (see Section \ref{sec:AstroResults}). Once we include DM parameters inference deteriorates. Main driver is the introduction of the parameter $m_\mathrm{WDM}$. We can interpret this as follows. The WDM mass influences the Jeans mass as $M_\mathrm{Jeans} \propto \left(\Omega_\mathrm{m}h^{2}\right)^{1/2}\left(\frac{m_\mathrm{WDM}}{keV}\right)^{-4}M_{\odot}$ holds (\cite{Sitwell_2014}), where we assume DM to be WDM. The halo mass of a collapsing halo is related to the virial temperature after the collapse via $T_\mathrm{vir}=\frac{\mu m_\mathrm{p}V^{2}_\mathrm{c}}{2k_\mathrm{B}}$, with mean molecular weight $\mu$, proton mass $m_\mathrm{p}$, Boltzmann constant $k_\mathrm{B}$, and circular velocity $V_\mathrm{c}$ (\cite{Barkana_2001}). Both parameters set a threshold for early star formation. If the Jeans mass is a more stringent boundary than the minimum virial temperature the latter will have only little influence on the era of reionisation.

This effect becomes obvious in the scatter plots in Figure~\ref{fig:Results}. The parameter $T_\mathrm{vir}$ shows a fairly decent agreement between predictions and labels for most points. There is however a certain subset of points with high scatter which get randomly assigned to false values independent of their true value. This is likely caused by the aforementioned influence of the WDM mass to set another threshold for efficient star formation, as the majority of the falsely attributed points stems from simulations with low WDM masses.

\begin{figure*}
\includegraphics[width=0.45\textwidth]{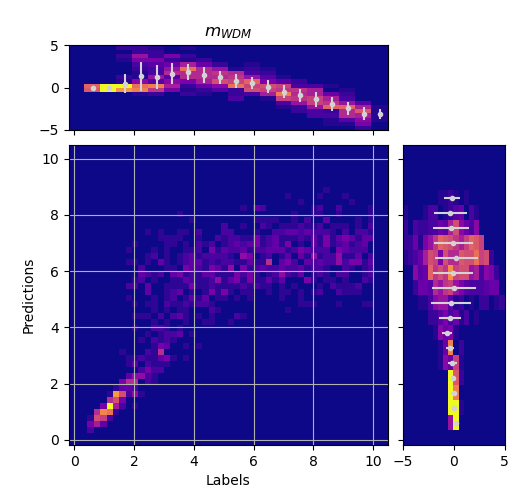}
\includegraphics[width=0.513\textwidth]{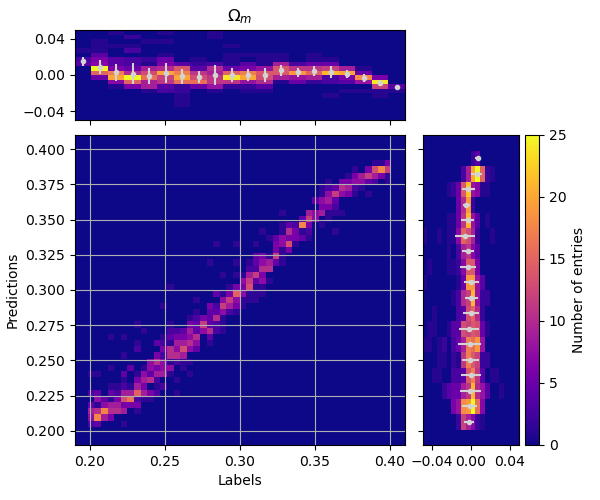}
\caption{Scatter plots for a detailed look at the DM parameters $m_\mathrm{WDM}$ (left) and $\Omega_\mathrm{m}$ (right). Both panels are accompanied by two histograms that show $y_{\rm i,pred}-y_{\rm i,true}$ at the position of $y_{\rm i,true}$ at the top and $y_{\rm i,true}-y_{\rm i,pred}$ at the position of $y_{\rm i,pred}$ at the right. The top and right histograms also depict the $1\sigma$ range for each bin. The colorbar applies to the top and right histograms. For the colorscale of the scatter plots we refer to Figure~\ref{fig:Results}.}
\label{fig:Results_WDM_OMm}
\end{figure*}

Finally we take a closer look at our two additional DM parameters $m_\mathrm{WDM}$ and $\Omega_\mathrm{m}$. Scatter plots of predicted versus true values for both parameters are displayed in Figure~\ref{fig:Results_WDM_OMm}. The extra top and right panels show the difference between prediction and true label given the true parameter label (top) and given the predicted parameter label (right); errorbars depict the 1$\sigma$ standard deviation. For $\Omega_\mathrm{m}$ the scatter plot showcases accurate inference across the chosen parameter range, away from the (s-shaped) prior boundaries. The scatter plot for $m_\mathrm{WDM}$ shows excellent agreement between true and predicted values for $m_\mathrm{WDM}<2 \,$keV, and the majority of predictions up to $m_\mathrm{WDM}=4 \,$keV is fairly accurate. However for labels in the $m_\mathrm{WDM}=2-4 \,$keV range outliers start to appear which scatter upwards to predict too high $m_\mathrm{WDM}$ values. For even higher masses above $m_\mathrm{WDM}> 4 \,$keV the predicting power of the network deteriorates. We can conclude that the range of WDM masses that can be well constrained via direct inference from 21cm tomographic light-cones strongly depends on the WDM mass and is restricted to lower masses below $\sim 4 \,$keV, as has been observed as well for summary statistics such as power spectra and bispectra~\citep{Carucci_2015,Saxena_2020}. Reason is that higher-mass WDM behaves more and more like CDM, with its power "cutoff" moving to smaller and smaller scales (e.g. \citet{Villanueva_Domingo_2018}). We note that future tomographic 21cm measurements during reionisation and cosmic dawn have the power to support current constraints e.g. from Ly$\alpha$ forest measurements that exclude low $m_\mathrm{WDM}$ masses. When solely training on $m_\mathrm{WDM}$ while keeping all other parameters constant we achieve $R^2>0.99$ on bare simulations, hinting at how well in principle $m_\mathrm{WDM}$ could be inferred by network-models when knowing the other model parameters perfectly well. Generally, inference from 21cm reionisation-era light-cones can be seen as a powerful additional tool to complement constraints on WDM.

\begin{figure}
\includegraphics[width=0.45\textwidth]{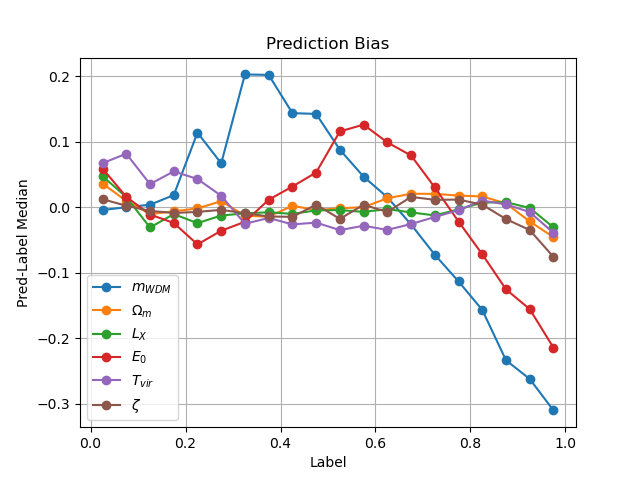}
\caption{Bias (solid lines) in the recovered parameter values in bins of ground truth parameter labels. The bias is calculated as the median difference $y_{\rm i,pred}-y_{\rm i,true}$ at the position of $y_{\rm i,true}$. Colours show the results for different parameters as indicated; parameter ranges are normalised to [0,1].
}
\label{fig:bias}
\end{figure}

As discussed above, a small bias arises towards the edges of the prior parameter ranges due to their sharp boundaries. Additionally, larger biases can arise for example due to models becoming indistinguishable above a certain threshold. Knowing the ground truth labels upon evaluation of the test set, we can directly measure bias present in our method as the median deviation between prediction and label. These trends are summarised in Figure~\ref{fig:bias} which reports the bias as a function of true parameter labels (labels are normalised between [0,1]). We note that the bias is comparably small $<5\%$ for most parameters and ranges, with rising biases due to sharp prior range boundaries more confined towards the edges, as also noticeable in Figure~\ref{fig:Results} and~\ref{fig:Results_WDM_OMm}. Only for $m_{\rm WDM}$ and $E_0$ the insensitivity of the signal to higher values and/or degeneracies hypothesised lead to the range where both parameters are unbiased being restricted mostly to the lower 20$\%$ of the allowed parameter range, translating to below $\sim 2\,$keV and $\sim 300\,$eV, respectively. For the other parameters this method is largely unbiased for most of the parameter range investigated. Lastly, there is possible bias stemming from the way the parameter set was sampled. As we randomly sampled 1000 parameter sets for the test set (4000 for the training set) we made sure the model predictive performance is not biased due to parameter sampling, with random per-parameter distances being considerably smaller than observed scatter between label and predicted parameter values.

%%%%%%%%%%%%%%%%%%%%%%%%%%%
%%%%%%%%%%%%%%%%%%%%%%%%%%%

%%%%%%%%%%%%%%%%%%%%%%%%%%%
%%%%%%%%%%%%%%%%%%%%%%%%%%%

%%%%%%%%%%%%%%%%%%%%%%%%%%%
%%%%%%%%%%%%%%%%%%%%%%%%%%%

%%%%%%%%%%%%%%%%%%%%%%%%%%%
\subsection{Parameter Inference from 21cm mock light-cones: Astrophysics and WDM Properties}\label{sec:MockResults}
%%%%%%%%%%%%%%%%%%%%%%%%%%%
Here we discuss our findings from training our 3D CNN on mock observed light-cones. The results for training and inferring on mock light-cones with optimal foreground settings (opt mock light-cones) are shown in row E in Table~\ref{tab:Results}. The difference between the $R^2$ and loss values for the opt mock and bare simulated light-cones is negligible.
Results for training our 3D CNN on mock light-cones with moderate foreground settings (mod mock light-cones) are shown in row F in Table~\ref{tab:Results}. Analogue to opt mock light-cones, mod mock light-cones are trained for 30 epochs. We stress here that the inference results for the mod mock light-cones only slightly deteriorate as compared to the performance of the 3D CNN on simulations or opt mock light-cones. In particular the parameters $\Omega_\mathrm{m}$ and $L_\mathrm{X}$ are barely affected by increased levels of foregrounds.

The inlet histograms in Figure~\ref{fig:Results} compare the difference between predictions and true parameter labels for opt mock light-cones (blue), mod mock light-cones (green) and bare simulated light-cones (red); they visualise the difference in accuracy and precision of inference for each parameter in these three cases. For $\Omega_\mathrm{m}$ and $L_\mathrm{X}$ inference for opt mock light-cones, mod mock light-cones and bare simulations works similarly well, with all three histograms centered around zero with narrow spread, i.e. indicating high accuracy and precision. For $T_\mathrm{vir}$ and $\zeta$ histograms indicate similarly low bias and scatter, with only a slightly larger scatter in the mod mock case. In the case of $E_0$ the histograms broaden, indicating lower precision achievable, again only with slightly increased scatter in the presence of noise and foregrounds. $m_\mathrm{WDM}$ is special insofar, as the predictions are unaffected for opt mock light-cones but become uninformative for mod mock light-cones. We therefore conclude that active foreground cleaning well into the wedge region is required to accurately infer $m_\mathrm{WDM}$ at low masses $\lesssim 4$ keV, while large scatter is present for higher masses that make up the broadened histogram base.

%%%%%%%%%%%%%%%%%%%%%%%%%%%
\renewcommand{\arraystretch}{1.5}
\setlength{\tabcolsep}{7.5pt}
\begin{table}
    \centering
    \begin{tabular}{l|cccc|c}
         Network Model &$L_\mathrm{X}$&$E_0$&$T_\mathrm{vir}$&$\zeta$&loss\\ \hline
          A. 3D CNN$_\mathrm{Sim}$&0.988&0.829&0.996&0.975&0.00436
         \\B. 3D CNN$_\mathrm{MA}$&0.985&0.843&0.993&0.976&0.00415
         \\C. LSTM$_\mathrm{Sim}$&0.993&0.867&0.999&0.989&0.00309
         \\D. LSTM$_\mathrm{MA}$&0.992&0.836&0.999&0.985&0.00387
    \end{tabular}
    \caption{Overview of $R^{2}$ and test loss values from training on the astro-only dataset (see Section~\ref{sec:AstroResults}). Results for the 3D CNN on astro-only (A.) and on mean averaged astro-only simulations (B.) are shown alongside results of the LSTM on astro-only (C.) and on mean averaged astro-only simulations (D.).}
    \label{tab:Astro_Results}
\end{table}
%%%%%%%%%%%%%%%%%%%%%%%%%%%

%%%%%%%%%%%%%%%%%%%%%%%%%%%
\subsection{Inference of Astrophysical Parameters only}
\label{sec:AstroResults}
%%%%%%%%%%%%%%%%%%%%%%%%%%%

Here we briefly present for comparison our findings from training our 3D CNN architecture exclusively on astrophysical parameters. The astro-only dataset (see Section~\ref{sec:datasets}) assumes a CDM universe and only varies the four astrophysical parameters $L_\mathrm{X}$, $E_0$, $T_\mathrm{vir}$ and $\zeta$.
The $R^2$ values inferred by our 3D CNN and LSTM on the astro-only test set are shown in Table~\ref{tab:Astro_Results}. In comparison to the results from training our 3D CNN on the full parameter set, see row D in Table~\ref{tab:Results}, and with the exception of $T_\mathrm{vir}$, almost the same $R^2$ values are reached for the astro-only parameters. For a discussion of the impact of degenerate behaviour with $m_\mathrm{WDM}$ for the inference of $T_\mathrm{vir}$ see Section~\ref{sec:3D_Results}. Note as well, that for the 3D CNN training and inference on a mean averaged set of light-cones (since interferometers are not capable of measuring the total brightness temperature but only relative brightness temperatures) had little to no effect on results (see row B in Table~\ref{tab:Astro_Results}) and yields performance similar to networks presented in earlier works~\citep{Gillet_2019,prelogovic2021machine}. Both on astro-only bare simulations and on the mean averaged sample the 3D CNN and LSTM perform very similar.
However, the 3D CNN gains an edge when working on more complex problems like training on the full astro+cosmo parameter set (see row C. in Table~\ref{tab:Results}). We conclude that our 3D CNN performs well in comparison to other well-optimised network types for parameter inference from 21 cm light-cones within a short training time and for a similar dataset.

A direct comparison of inference results obtained via Markov Chain Monte Carlo (MCMC) analysis of 3D 21cm light-cones (see~\citet{2018Greig} using 21CMMC\footnote{https://github.com/BradGreig/21CMMC}) to the 3D CNN parameter recovery presented in this work is difficult. Marginalised uncertainties needed to be compared with the scatter in recovered parameters. Furthermore, assumptions on the fiducial reionisation model and foregrounds differ. Extending our network to variational or Bayesian inference, e.g.~\citet{hortua2020parameters}, could enable a more direct comparison to marginalized uncertainties. We leave this extension of our 3D CNN for future work.
Nevertheless, to compare our network performance with an MCMC analysis, we estimate the uncertainty of the 3D CNN on the "faint" galaxy model used in~\citet{2018Greig}. To this end, we calculated the standard deviation of differences between predictions and labels for a parameter region restricted to this "faint" galaxy model $\pm 5\%$ of our total parameter range defined in Section~\ref{sec:parameters}. The resulting deviations are $\sigma_{\mathrm{log}(L_\mathrm{X})=40}=0.084$, $\sigma_{E_0=500 eV}=78$ eV, $\sigma_{\mathrm{log}(T_\mathrm{vir})=4.7}=0.022$, $\sigma_{\zeta=30}=3.9$. Note that we did not restrict all parameters simultaneously. The scatter derived is competitive with MCMC methods applied to one (mock) 21cm observation at the time. We stress the advantage of our network-based parameter recovery with a 3D CNN being able to simultaneously explore the entire parameter space of interest, while MCMC-based predictions are restricted to one fiducial model. At the same time constraints are not biased due to the chosen (often Gaussian) summary statistics and use the full range of information available in the training set, potentially yielding more accurate results than the MCMC analysis as shown by~\citet{zhao2021simulationbased}.

\renewcommand{\arraystretch}{1.5}
\setlength{\tabcolsep}{4.5pt}
\begin{table}
    \centering
    \begin{tabular}{l|cccccc}
         Network Model &$m_\mathrm{WDM}$&$\Omega_\mathrm{m}$&$L_\mathrm{X}$&$E_0$&$T_\mathrm{vir}$&$\zeta$\\ \hline
         3D CNN$_\mathrm{Mock\rightarrow Sim}$&-0.759&0.024&\bf{0.969}&\bf{0.768}&-1.21&\bf{0.930}
         \\3D CNN$_\mathrm{Sim\rightarrow Mock}$&-0.329&\bf{0.748}&\bf{0.981}&\bf{0.809}&0.115&\bf{0.935}
    \end{tabular}
    \caption{$R^2$ values from transfer learning as described in Section~\ref{sec:TransferLearning}. Results for training our 3D CNN on opt mock light-cones and then inferring parameter values from bare simulations (3D CNN$_\mathrm{Mock\rightarrow Sim}$) and for training our 3D CNN on bare simulations and then inferring parameter values from opt mock light-cones (3D CNN$_\mathrm{Sim\rightarrow Mock}$) are shown. In bold, we highlight where training on simulations followed by inference from mock light-cones (and vice-versa) is able to conclusively predict parameters.}
    \label{tab:Transfer_Learning}
\end{table}
%%%%%%%%%%%%%%%%%%%%%%%%%%%

%%%%%%%%%%%%%%%%%%%%%%%%%%%
\begin{figure}
\includegraphics[width=0.45\textwidth]{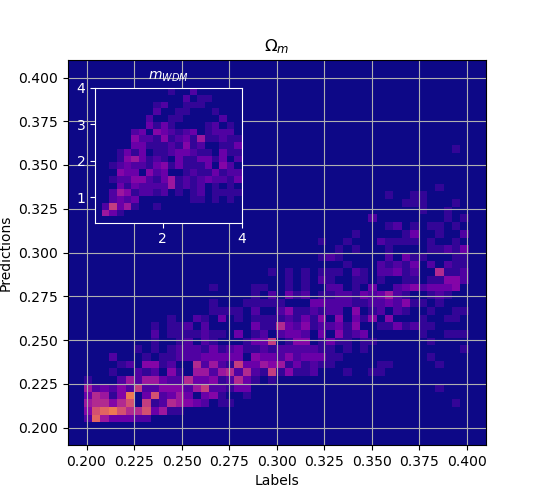}
\includegraphics[width=0.548\textwidth]{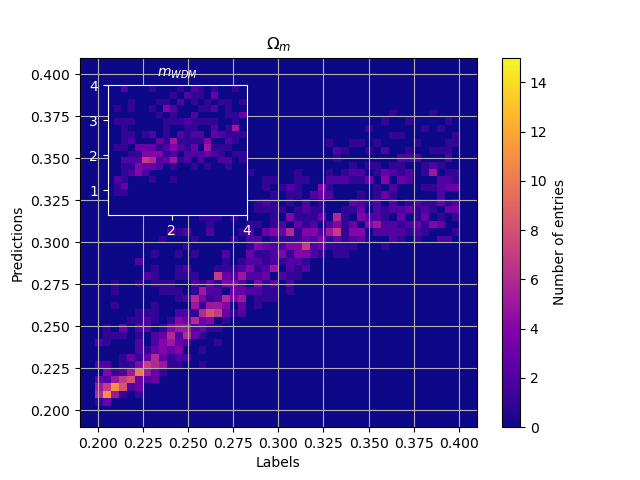}
\caption{Scatter plots of predicted versus true parameter values (labels) for transfer learning, of WDM parameters according to Section~\ref{sec:TransferLearning}. Left:  Predictions of the 3D CNN for simulated light-cones after being trained on opt mock light-cones. Right: Predictions of the 3D CNN for opt mock light-cones after being trained on simulations. The inlets show the scatter of predictions versus labels for $m_\mathrm{WDM}< 4$ keV. Note that the equivalent scatter plots for the remaining parameters are shown in Figure~\ref{fig:TransferLearningAppendix} in the appendix. }
\label{fig:TransferLearning}
\end{figure}
%%%%%%%%%%%%%%%%%%%%%%%%%%%

%%%%%%%%%%%%%%%%%%%%%%%%%%%
\subsection{Transfer Learning: Inference from Cross Training}
\label{sec:TransferLearning}
%%%%%%%%%%%%%%%%%%%%%%%%%%%

\noindent For 21cm measurements during the EoR and Cosmic Dawn accurate modelling of interferometer noise and especially of foregrounds many orders of magnitude above the expected signal remain a challenging task. In this section we want to test the robustness of our 3D CNN-derived parameter predictions against inaccuracies in noise and foreground modelling. We therefore take the approach to cross train, or transfer learn, between bare simulated and mock light-cones that include foreground and noise. We do so by first training our 3D CNN on bare simulated light-cones and then infer parameters from opt mock light-cones and vice versa. We thus reuse the respective pre-trained model, freeze the model, and use it for parameter inference without any re-training. This analysis helps us understand the maximum impact training on a database that mis-specified the amount of noise and foregrounds could have on the network's inference performance. The required computational time for this analysis is thus only the network evaluation time (seconds on one GPU -- about 3 seconds per batch of eight observed instances for the 3D CNN). Note that this is a fairly conservative scenario as it, for example, tests the ability of parameter inference from noisy data when knowing before only simulations. In reality the difference between a training set and data taken is expected to be significantly smaller. 

The $R^2$ values for parameter inference for our set of astrophysical and DM parameters are shown in Table~\ref{tab:Transfer_Learning}, for the case of transfer learning after training on simulations to mock light-cones (3D CNN$_\mathrm{Sim\rightarrow Mock}$), and inversely for transfer learning after training on mock light-cones to simulated light-cones (3D CNN$_\mathrm{Mock\rightarrow Sim}$). In both cases we get excellent results for inference of the X-ray luminosity $L_\mathrm{X}$ and ionising efficiency $\zeta$, as well as good results for the X-ray energy threshold $E_0$, presumably driven mostly by timing and shape of fluctuations towards the end of reionisation where the difference between opt mock light-cones and simulations is small. Also, training on simulations and inference from mock light-cones yields reasonable results for $\Omega_\mathrm{m}$, which is not true in the opposite case. Some potential reasons will briefly be discussed in Section~\ref{sec:saliency}, where we take a look at the network's attention. For the two degenerate and harder-to-infer parameters $m_\mathrm{WDM}$ and $T_\mathrm{vir}$ in neither case meaningful inference is possible. We conclude that both parameters require accurate noise and foreground modelling for inference.
Figure~\ref{fig:TransferLearning} shows the scatter plots for predictions against true values for WDM mass $m_\mathrm{WDM}$ (inlets) and DM density $\Omega_\mathrm{m}$ (large panels), each for predictions on simulations after training on mock light-cones (left) and vice versa (right). For the remaining astrophysical parameters we refer to Figure~\ref{fig:TransferLearningAppendix} in the appendix. Predictions on mock light-cones of $\Omega_\mathrm{m}$ (right) yield reasonable results, but with increased scatter, while the inverse case (left) is biased towards low values. For the WDM mass the inlets zoom in at $m_\mathrm{WDM} < 4\,$keV, where we note significantly increased scatter as compared to previous inference without transfer.

%%%%%%%%%%%%%%%%%%%%%%%%%%%
\begin{figure*}
\includegraphics[width=\textwidth]{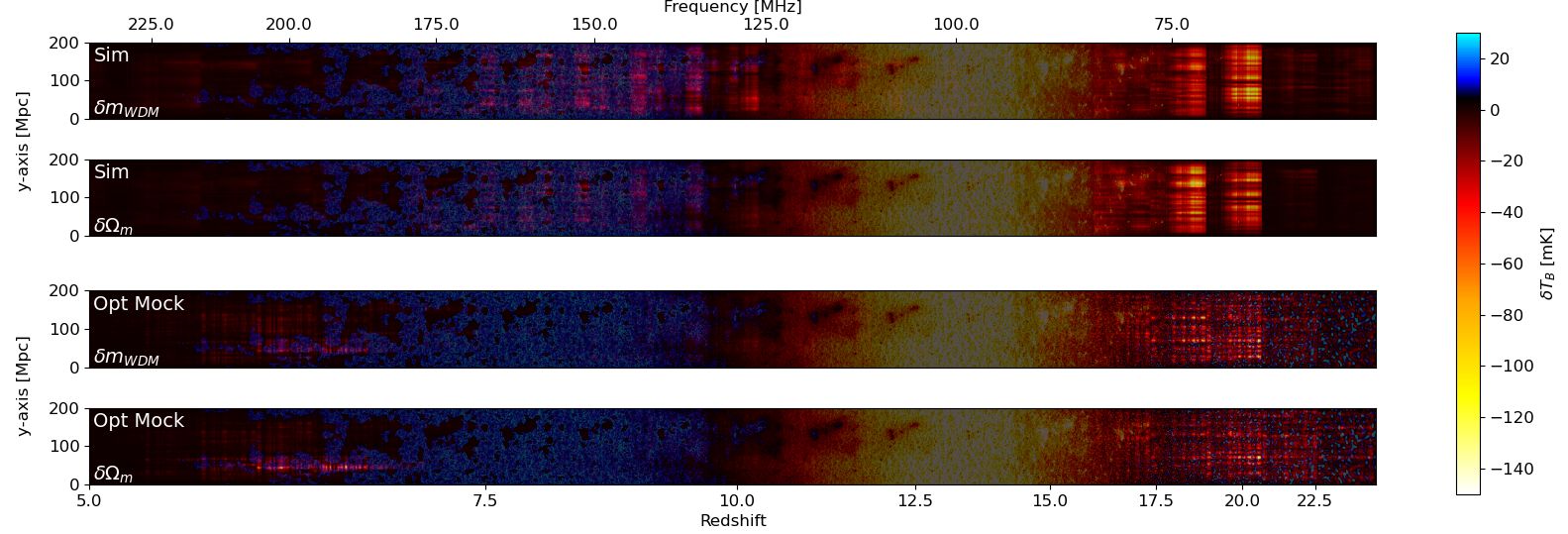}
\caption{Light-cone slice overlaid with saliency maps (red to yellow for highest saliency) for our 3D CNN after being trained on simulations (top two panels) and opt mock light-cones (bottom two panels), with respective top slices depicting saliency for the WDM mass $m_\mathrm{WDM}$ and bottom slices depicting saliency for the DM density $\Omega_\mathrm{m}$ as indicated in white on each slice. The colorbar on the right denotes the 21cm brightness temperature fluctuations of the light-cone slice.}
\label{fig:Saliency}
\end{figure*}
%%%%%%%%%%%%%%%%%%%%%%%%%%%

%%%%%%%%%%%%%%%%%%%%%%%%%%%
\section{3D CNN Network Interpretation} \label{sec:interpretation}
%%%%%%%%%%%%%%%%%%%%%%%%%%%
In order to understand which underlying features guide the best-performing 3D CNN for parameter inference from 21cm light-cones, this section takes a closer look at saliency maps, with focus on our two DM parameters, as well as filter structure.

%%%%%%%%%%%%%%%%%%%%%%%%%%%
\subsection{Saliency Maps} \label{sec:saliency}
%%%%%%%%%%%%%%%%%%%%%%%%%%%

\noindent Gradient-based saliency maps, first proposed by \citet{simonyan2014deep}, represent the gradient of the output with respect to the input values. The resulting saliency maps highlight regions that the network focuses on, where small changes yield large effects on the output. For creation of the saliency maps we used the publicly available code \textit{tf-keras-vis}~\citep{keisentfkerasvis}.

We show fiducial light-cone slices overlaid with saliency maps for our 3D CNN after being trained on simulations as well as on opt mock light-cones in Figure~\ref{fig:Saliency}. We here focus on the DM parameters $m_\mathrm{WDM}$ and $\Omega_\mathrm{m}$. For saliency maps of the remaining parameters we refer to Figure~\ref{fig:SaliencyAppendix} in the appendix. The light-cone slice was created for fixed $m_\mathrm{WDM}=2\,$keV (a mass that our network is capable of getting a good prediction for), as well as fiducial parameters $\Omega_\mathrm{m}=0.31$, log$(L_\mathrm{X})=40.0$, $E_0=500\,$eV, log$(T_\mathrm{vir})=4.7$ and $\zeta=30$. For each saliency map four light-cones of the fiducial parameter set and with different random seeds were created. Then the saliency map for each light-cone is derived and a 2D slice from each of the four saliency maps taken. The average of the absolute value of the saliency is plotted over one of the four original light-cone slices. Note the saliency maps are normalised between zero and their respective maxima. The normalisation for the mock saliency maps is 1.5-3 times higher than for the simulated light-cones. 

When comparing the saliency for simulated and mock light-cones in Figure~\ref{fig:Saliency}, in the case of mock light-cones the CNN generally focuses more strongly on smaller, well defined areas, while for simulations attention is more broadly distributed across the full light-cone. This might partly explain the asymmetry in transfer learning performance as shown in Section~\ref{sec:TransferLearning}. In both cases information from high redshifts at the onset of heating and low redshifts towards the end of reionisation is important for parameter inference, even for high levels of noise at high redshifts in the case of the mock light-cones. We note that this effect of focus on 'transition' regions e.g. between heating and reionisation is pronounced as well in the case of mod mock light-cones that display higher levels of noise than opt mock light-cones. That high redshifts hold significant information is, for example expected for low $m_\mathrm{WDM}$, as it delays early structure formation. We stress that parameter results deteriorate when not taking into account high-redshift (heating era) fluctuations for inference. This might explain why for the transfer learning test of training on mock light-cones and inference from simulations (see \ref{sec:TransferLearning}) $\Omega_\mathrm{m}$ yields biased results with the absence of high-redshift noise mistaken as delayed structure formation by the network. In contrast, after training on simulations the network expects no noise at high redshifts, but still manages to correctly identify the onset of the early heating signal in mock light-cones.

%%%%%%%%%%%%%%%%%%%%%%%%%%%
\subsection{Filter Structures}
\label{sec:Filter}
%%%%%%%%%%%%%%%%%%%%%%%%%%%
The first layer of our neural network applied to input light-cones has a kernel size of 3x3x102, and 32 filters. The large kernel size in frequency (redshift) direction allows for global structures in the weight values of most filters. We show a representative set of four filters in the appendix in Figure~\ref{fig:Filter}, where weights were averaged over the two spatial dimensions of each filter to be plotted against the third frequency dimension. We generally identify three types of filters, mountains and troughs that we attribute the bubble-like structures and that are by far the most common filters, rises and falls that seem to map general trends in brightness temperature, and more complicated unidentifiable structures. We note as well that the focus on 'transition areas' with rising and falling brightness as well as on larger-scale fluctuations matches the areas identified as the main focus of the saliency maps, e.g. towards the end of reionisation, in the previous section.

%%%%%%%%%%%%%%%%%%%%%%%%%%%
\section{Summary and Conclusions}
\label{sec:conclusions}
%%%%%%%%%%%%%%%%%%%%%%%%%%%
Radio interferometers such as the SKA will allow for precise measurements of the 21 cm signal at high redshifts of reionisation and beyond. Taking full advantage in the analysis of the highly non-Gaussian information encoded in the 21 cm signal may be challenging with traditional summary statistics like the power spectrum. 

In this paper, we employ neural networks to jointly infer astrophysics and DM properties directly from 21 cm light-cones. These parameters are the X-ray luminosity $L_\mathrm{X}$, the ionising efficiency $\zeta$, the minimum virial temperature for star forming haloes, $T_\mathrm{vir}$, and the X-ray energy threshold, $E_0$, as well as for DM the matter density, $\Omega_\mathrm{m}$, and the WDM mass, $m_\mathrm{WDM}$. For this purpose we produced a dataset of 5050 light-cone simulations. To find an optimal architecture tailored for such tomographic light-cone data we tested ResNet, LSTM as well 2D and 3D CNN architectures. After thorough hyperparameter optimisation for the models we found a simple 3D CNN architecture to perform best. The first convolutional layer of the 3D CNN has a relatively large kernel size in frequency direction to enable the network to follow the redshift-dependent global evolution of the 21cm signal, while being trainable with low computational cost. An additional GAP layer was found to significantly outperform alternative methods against overfitting with regards to smooth and quick convergence.

The model is able to infer astrophysical and DM parameters simultaneously and accurately. For parameter inference from simulations, the 3D CNN achieves an excellent performance for the X-ray luminosity $L_\mathrm{X}$, matter density $\Omega_\mathrm{m}$ and ionising efficiency $\zeta$ with coefficients of determination $R^2 > 0.97$. The virial temperature, $T_\mathrm{vir}$, and the X-ray energy threshold, $E_0$, proved to be slightly more challenging with $R^2 > 0.82$ and $R^2 > 0.77$ respectively. For $T_\mathrm{vir}$ we attribute the loss in performance compared to a dataset of astrophysical parameters alone to a degeneracy with $m_\mathrm{WDM}$. Especially for low WDM masses this puts a more stringent threshold on star formation. We confirmed this suspicion by training on an astrophysics-only dataset (with DM parameters fixed to fiducial values), where we found most parameters to be unaffected by the inclusion of the DM parameters, but for $T_\mathrm{vir}$ to then reach very high $R^2 > 0.99$. 
The energy threshold $E_0$ is inferred similarly well, regardless of an astrophysics-only set of parameters or joint constraints with DM which corresponds to findings in previous works. In case of the mass of WDM, $m_\mathrm{WDM}$, precise and accurate results for $m_\mathrm{WDM} < 2\,$keV are achieved. For the mass range $m_\mathrm{WDM}=2-4\,$keV the predictions of the 3D CNN show increased scatter and first outliers of high-mass predictions. For values above $m_\mathrm{WDM}\approx 4\,$keV the prediction capability of our 3D CNN stalls. Through their typical free-streaming length and corresponding small-scale modifications of the dark matter power spectrum underlying the 21 cm signal, one can thus test thermal relics at few keV that are semi-relativistic and decoupled well before ordinary neutrinos; prominent examples being sterile neutrinos and the gravitino.

We find that for mock light-cones in an optimistic foreground scenario inference performance is unaffected as compared to bare simulations, while even in a more moderate foreground scenario the results only slightly diminish. The parameters $\Omega_\mathrm{m}$, $L_\mathrm{X}$ and $\zeta$ have been found to be the most resilient against noise and foreground contamination and performed well in all three cases (simulations only, mock light-cones with optimistic foregrounds, mock light-cones with moderate foreground levels). All three parameters can strongly affect the EoR, where noise is lower.
As interferometers are not capable of measuring absolute brightness temperature values but only relative ones, we tested the influence of mean averaging on the performance, which proved to work equally well for an astrophysics-only dataset to inference from bare simulations. Compared to previous work~\citep{Gillet_2019,zhao2021simulationbased,prelogovic2021machine,2021FDM_CNN} our 3D CNN performed comparably well on an astrophysics-only dataset, with fast training times on a moderately small dataset, while at the same time being able to robustly infer a larger set of parameters, in particular DM parameters jointly with astrophysics.

As for the 21cm signal during the EoR and Cosmic Dawn the modelling of noise and especially foregrounds many orders of magnitude higher than the cosmological signal comes with uncertainties, we test how inaccuracies in our training set affect our results with transfer learning between bare simulated and mock light-cones (by training our 3D CNN on bare simulations and using it to predict on mock light-cones and vice versa). In both cases we get excellent results for $L_\mathrm{X}$, $E_0$ and $\zeta$ as compared to the standard case. In case of inference from mock light-cones after training on bare simulations we get robust results for the parameter $\Omega_\mathrm{m}$. The remaining parameters $m_\mathrm{WDM}$ and $T_\mathrm{vir}$ cannot be robustly inferred after cross-training; they therefore should be regarded as more sensitive towards inaccuracies in noise and foreground modelling.

Finally, we produced saliency maps for a better understanding of our 3D CNN's attention. The saliency maps highlight regions of the light-cones important for inference of respective parameters. These show differences in the case of training on opt mock light-cones and bare simulations. For bare simulations, high saliency areas are spread across larger stretches of the light-cones in frequency (redshift) which may explain the aforementioned performance discrepancies in transfer learning. It is interesting to note that in both cases the 21cm brightness fluctuation signal stemming from the onset of heating and the last stages of reionisation were especially important for inference of both astrophysical and DM parameters according to the saliency signal, further stressing the importance and constraining power of both CD and EoR measurements of the cosmological 21cm signal.

A more direct comparison of parameter recovery from 21cm light-cones based on the 3D CNN presented in this work with inference results obtained for example with an MCMC analysis warrants an extension of our network to a variational or Bayesian approach. Here our work provides point estimates alongside scatter and bias for a wide range of models, and also investigates the impact of mis-specified noise and foregrounds via transfer modelling. We would like to stress, that a well-optimised architecture that pushes for finding an as good as possible model (network) fit with small scatter is needed in order to move on to well-motivated uncertainty calibration. For e.g. a Bayesian network this can entail to calibrate its uncertainty estimates against our scatter of point estimates~\citep{2022arXiv220101203M}. We caution though that as well-calibrated error estimates have not been a focus of deep learning and are becoming important now through our scientific applications, more work is needed for reliable and robust estimates.  Given our good network performance, while avoiding overfitting and being relatively small and fast to train, makes our 3D-PieNet a good candidate for such a calibration.

Advantages of our network-based approach firstly are a simultaneous exploration of the entire parameter space of interest; secondly, inferred parameter values are not biased due to the choice of statistics, avoiding biases, while learning representations present in the training 21 cm light-cones. Furthermore, these methods can be complementary to 'traditional' methods to better understand model and data properties, using for example saliency mappings or transfer learning. Also, besides the mentioned ability to pick up e.g. beyond Gaussian information, computational speed can be gained, especially interesting for further expanded parameter sets. Even the need to re-train with new data (in our case taking about a day on one GPU) is small as compared to the cost of MCMC'ing 21cm lightcones. This makes networks good emulators, or at least exploratory tools in the future.
New ways to tackle for example the combination of constraints from different probes might be waiting around the corner, such as instead of using posteriors to combine constraints, one could work directly with multi-channel inputs for network models, e.g. inputting a CMB map alongside a 21 cm lightcone or other LSS probes. We thus advocate for deep learning techniques to supplement and expand our astrophysical and cosmological toolbox.

%%%%%%%%%%%%%%%%%%%%%%%%%%%
\section*{Acknowledgements}
%%%%%%%%%%%%%%%%%%%%%%%%%%%
We thank the anonymous referee for the useful comments that
improved this work. We would like to thank Gregor Kasieczka for his idea of using a convolutional layer with a large kernel size in redshift direction, which ended up being a key component in our 3D CNN. This work is funded by the Deutsche Forschungsgemeinschaft (DFG, German Research Foundation) under Germany's Excellence Strategy -- EXC 2121 ``Quantum Universe'' --  390833306.

%%%%%%%%%%%%%%%%%%%%%%%%%%%
\section*{Data Availability}
%%%%%%%%%%%%%%%%%%%%%%%%%%%
The code, trained network weights and data underlying this paper are available at https://github.com/stef-neu/3D-21cmPIE-Net.

%%%%%%%%%%%%%%%%%%%%%%%%%%%
%%%%%%%%%%%%%%%%%%%%%%%%%%%
\bibliographystyle{mnras}
\bibliography{mybib}

\providecommand{\noopsort}[1]{}\providecommand{\singleletter}[1]{#1}%
\begin{thebibliography}{}
\makeatletter
\relax
\def\mn@urlcharsother{\let\do\@makeother \do\$\do\&\do\#\do\^\do\_\do\%\do\~}
\def\mn@doi{\begingroup\mn@urlcharsother \@ifnextchar [ {\mn@doi@}
  {\mn@doi@[]}}
\def\mn@doi@[#1]#2{\def\@tempa{#1}\ifx\@tempa\@empty \href
  {http://dx.doi.org/#2} {doi:#2}\else \href {http://dx.doi.org/#2} {#1}\fi
  \endgroup}
\def\mn@eprint#1#2{\mn@eprint@#1:#2::\@nil}
\def\mn@eprint@arXiv#1{\href {http://arxiv.org/abs/#1} {{\tt arXiv:#1}}}
\def\mn@eprint@dblp#1{\href {http://dblp.uni-trier.de/rec/bibtex/#1.xml}
  {dblp:#1}}
\def\mn@eprint@#1:#2:#3:#4\@nil{\def\@tempa {#1}\def\@tempb {#2}\def\@tempc
  {#3}\ifx \@tempc \@empty \let \@tempc \@tempb \let \@tempb \@tempa \fi \ifx
  \@tempb \@empty \def\@tempb {arXiv}\fi \@ifundefined
  {mn@eprint@\@tempb}{\@tempb:\@tempc}{\expandafter \expandafter \csname
  mn@eprint@\@tempb\endcsname \expandafter{\@tempc}}}

\bibitem[\protect\citeauthoryear{Abazajian, Fuller  \& Patel}{Abazajian
  et~al.}{2001}]{Abazajian_2001}
Abazajian K.,  Fuller G.~M.,   Patel M.,  2001, \mn@doi [Physical Review D]
  {10.1103/physrevd.64.023501}, 64

\bibitem[\protect\citeauthoryear{Adhikari et~al.,}{Adhikari
  et~al.}{2017}]{Adhikari_2017}
Adhikari R.,  et~al., 2017, \mn@doi [Journal of Cosmology and Astroparticle
  Physics] {10.1088/1475-7516/2017/01/025}, 2017, 025

\bibitem[\protect\citeauthoryear{Aghanim et~al.,}{Aghanim
  et~al.}{2020}]{Planck2018}
Aghanim N.,  et~al., 2020, \mn@doi [Astronomy \& Astrophysics]
  {10.1051/0004-6361/201833910}, 641, A6

\bibitem[\protect\citeauthoryear{Barkana \& Loeb}{Barkana \&
  Loeb}{2001}]{Barkana_2001}
Barkana R.,  Loeb A.,  2001, \mn@doi [Physics Reports]
  {10.1016/s0370-1573(01)00019-9}, 349, 125–238

\bibitem[\protect\citeauthoryear{{Berti}, {Spinelli}, {Haridasu}, {Viel}  \&
  {Silvestri}}{{Berti} et~al.}{2021}]{Berti2021}
{Berti} M.,  {Spinelli} M.,  {Haridasu} B.~S.,  {Viel} M.,   {Silvestri} A.,
  2021, arXiv e-prints, \href
  {https://ui.adsabs.harvard.edu/abs/2021arXiv210903256B} {p. arXiv:2109.03256}

\bibitem[\protect\citeauthoryear{{Bond} \& {Szalay}}{{Bond} \&
  {Szalay}}{1983}]{WDM1983}
{Bond} J.~R.,  {Szalay} A.~S.,  1983, \mn@doi [Astrophys. J.] {10.1086/161460},
  \href {https://ui.adsabs.harvard.edu/abs/1983ApJ...274..443B} {274, 443}

\bibitem[\protect\citeauthoryear{Boyarsky, Lesgourgues, Ruchayskiy  \&
  Viel}{Boyarsky et~al.}{2009}]{PhysRevLett.102.201304}
Boyarsky A.,  Lesgourgues J.,  Ruchayskiy O.,   Viel M.,  2009, \mn@doi [Phys.
  Rev. Lett.] {10.1103/PhysRevLett.102.201304}, 102, 201304

\bibitem[\protect\citeauthoryear{Boylan-Kolchin, Bullock  \&
  Kaplinghat}{Boylan-Kolchin et~al.}{2011}]{Boylan_Kolchin_2011}
Boylan-Kolchin M.,  Bullock J.~S.,   Kaplinghat M.,  2011, \mn@doi [Monthly
  Notices of the Royal Astronomical Society: Letters]
  {10.1111/j.1745-3933.2011.01074.x}, 415, L40–L44

\bibitem[\protect\citeauthoryear{Carucci, Villaescusa-Navarro, Viel  \&
  Lapi}{Carucci et~al.}{2015}]{Carucci_2015}
Carucci I.~P.,  Villaescusa-Navarro F.,  Viel M.,   Lapi A.,  2015, \mn@doi
  [Journal of Cosmology and Astroparticle Physics]
  {10.1088/1475-7516/2015/07/047}, 2015, 047

\bibitem[\protect\citeauthoryear{{Das}, {Mesinger}, {Pallottini}, {Ferrara}  \&
  {Wise}}{{Das} et~al.}{2017}]{Das2017}
{Das} A.,  {Mesinger} A.,  {Pallottini} A.,  {Ferrara} A.,   {Wise} J.~H.,
  2017, \mn@doi [\mnras] {10.1093/mnras/stx943}, \href
  {https://ui.adsabs.harvard.edu/abs/2017MNRAS.469.1166D} {469, 1166}

\bibitem[\protect\citeauthoryear{DeBoer et~al.,}{DeBoer
  et~al.}{2017}]{DeBoer_2017}
DeBoer D.~R.,  et~al., 2017, \mn@doi [Publications of the Astronomical Society
  of the Pacific] {10.1088/1538-3873/129/974/045001}, 129, 045001

\bibitem[\protect\citeauthoryear{Dodelson \& Widrow}{Dodelson \&
  Widrow}{1994}]{Dodelson_1994}
Dodelson S.,  Widrow L.~M.,  1994, \mn@doi [Physical Review Letters]
  {10.1103/physrevlett.72.17}, 72, 17–20

\bibitem[\protect\citeauthoryear{Evoli, Mesinger  \& Ferrara}{Evoli
  et~al.}{2014}]{Evoli_2014}
Evoli C.,  Mesinger A.,   Ferrara A.,  2014, \mn@doi [Journal of Cosmology and
  Astroparticle Physics] {10.1088/1475-7516/2014/11/024}, 2014, 024–024

\bibitem[\protect\citeauthoryear{{Furlanetto}, {Zaldarriaga}  \&
  {Hernquist}}{{Furlanetto} et~al.}{2004}]{2004reionexcursion}
{Furlanetto} S.~R.,  {Zaldarriaga} M.,   {Hernquist} L.,  2004, \mn@doi [\apj]
  {10.1086/423028}, \href
  {https://ui.adsabs.harvard.edu/abs/2004ApJ...613...16F} {613, 16}

\bibitem[\protect\citeauthoryear{Garzilli, Magalich, Ruchayskiy  \&
  Boyarsky}{Garzilli et~al.}{2021}]{10.1093/mnras/stab192}
Garzilli A.,  Magalich A.,  Ruchayskiy O.,   Boyarsky A.,  2021, \mn@doi
  [Monthly Notices of the Royal Astronomical Society] {10.1093/mnras/stab192},
  502, 2356

\bibitem[\protect\citeauthoryear{Gillet, Mesinger, Greig, Liu  \& Ucci}{Gillet
  et~al.}{2019}]{Gillet_2019}
Gillet N.,  Mesinger A.,  Greig B.,  Liu A.,   Ucci G.,  2019, \mn@doi [Monthly
  Notices of the Royal Astronomical Society] {10.1093/mnras/stz010}

\bibitem[\protect\citeauthoryear{Greig \& Mesinger}{Greig \&
  Mesinger}{2015}]{Greig2015}
Greig B.,  Mesinger A.,  2015, \mn@doi [Monthly Notices of the Royal
  Astronomical Society] {10.1093/mnras/stv571}, 449, 4246–4263

\bibitem[\protect\citeauthoryear{Greig \& Mesinger}{Greig \&
  Mesinger}{2017}]{Greig2017}
Greig B.,  Mesinger A.,  2017, \mn@doi [Proceedings of the International
  Astronomical Union] {10.1017/s1743921317011103}, 12, 18–21

\bibitem[\protect\citeauthoryear{Greig \& Mesinger}{Greig \&
  Mesinger}{2018}]{2018Greig}
Greig B.,  Mesinger A.,  2018, \mn@doi [Monthly Notices of the Royal
  Astronomical Society] {10.1093/mnras/sty796}, 477, 3217–3229

\bibitem[\protect\citeauthoryear{{Hassan}, {Andrianomena}  \&
  {Doughty}}{{Hassan} et~al.}{2020}]{2020HassanCNN}
{Hassan} S.,  {Andrianomena} S.,   {Doughty} C.,  2020, \mn@doi [\mnras]
  {10.1093/mnras/staa1151}, \href
  {https://ui.adsabs.harvard.edu/abs/2020MNRAS.494.5761H} {494, 5761}

\bibitem[\protect\citeauthoryear{He, Zhang, Ren  \& Sun}{He
  et~al.}{2016}]{He2016DeepRecognition}
He K.,  Zhang X.,  Ren S.,   Sun J.,  2016, {Deep Residual Learning for Image
  Recognition}, \url {http://image-net.org/challenges/LSVRC/2015/}

\bibitem[\protect\citeauthoryear{Heneka \& Amendola}{Heneka \&
  Amendola}{2018a}]{Heneka:2018kgn}
Heneka C.,  Amendola L.,  2018a, in {53rd Rencontres de Moriond on Cosmology}.
  pp 207--210 (\mn@eprint {arXiv} {1805.11044})

\bibitem[\protect\citeauthoryear{{Heneka} \& {Amendola}}{{Heneka} \&
  {Amendola}}{2018b}]{Heneka2018}
{Heneka} C.,  {Amendola} L.,  2018b, \mn@doi [\jcap]
  {10.1088/1475-7516/2018/10/004}, \href
  {https://ui.adsabs.harvard.edu/abs/2018JCAP...10..004H} {2018, 004}

\bibitem[\protect\citeauthoryear{{Heneka} \& {Mesinger}}{{Heneka} \&
  {Mesinger}}{2020}]{2020MNRAS.496..581H}
{Heneka} C.,  {Mesinger} A.,  2020, \mn@doi [\mnras] {10.1093/mnras/staa1517},
  \href {https://ui.adsabs.harvard.edu/abs/2020MNRAS.496..581H} {496, 581}

\bibitem[\protect\citeauthoryear{Hochreiter \& Schmidhuber}{Hochreiter \&
  Schmidhuber}{1997}]{HochSchm97}
Hochreiter S.,  Schmidhuber J.,  1997, Neural Computation, 9, 1735

\bibitem[\protect\citeauthoryear{Hortúa, Volpi  \& Malagò}{Hortúa
  et~al.}{2020a}]{hortua2020parameters}
Hortúa H.~J.,  Volpi R.,   Malagò L.,  2020a, Parameters Estimation from the
  21 cm signal using Variational Inference (\mn@eprint {arXiv} {2005.02299})

\bibitem[\protect\citeauthoryear{Hortúa, Malagò  \& Volpi}{Hortúa
  et~al.}{2020b}]{Hort_a_2020}
Hortúa H.~J.,  Malagò L.,   Volpi R.,  2020b, \mn@doi [Machine Learning:
  Science and Technology] {10.1088/2632-2153/aba6f1}, 1, 035014

\bibitem[\protect\citeauthoryear{Ioffe \& Szegedy}{Ioffe \&
  Szegedy}{2015}]{ioffe2015batch}
Ioffe S.,  Szegedy C.,  2015, Batch Normalization: Accelerating Deep Network
  Training by Reducing Internal Covariate Shift (\mn@eprint {arXiv}
  {1502.03167})

\bibitem[\protect\citeauthoryear{Iqbal et~al.}{Iqbal
  et~al.}{2020}]{PlotNeuralNet}
Iqbal H.,  et~al., 2020, PlotNeuralNet,
  \url{https://github.com/HarisIqbal88/PlotNeuralNet}

\bibitem[\protect\citeauthoryear{Iršič et~al.,}{Iršič
  et~al.}{2017}]{Ir_i__2017}
Iršič V.,  et~al., 2017, \mn@doi [Physical Review D]
  {10.1103/physrevd.96.023522}, 96

\bibitem[\protect\citeauthoryear{{Jones}, {Palatnick}, {Chen}, {Beane}  \&
  {Lidz}}{{Jones} et~al.}{2021}]{FDM2021}
{Jones} D.,  {Palatnick} S.,  {Chen} R.,  {Beane} A.,   {Lidz} A.,  2021,
  \mn@doi [\apj] {10.3847/1538-4357/abf0a9}, \href
  {https://ui.adsabs.harvard.edu/abs/2021ApJ...913....7J} {913, 7}

\bibitem[\protect\citeauthoryear{Keisen et~al.}{Keisen
  et~al.}{2020}]{keisentfkerasvis}
Keisen et~al., 2020, tf-keras-vis, \url{https://github.com/keisen/tf-keras-vis}

\bibitem[\protect\citeauthoryear{Kern, Liu, Parsons, Mesinger  \& Greig}{Kern
  et~al.}{2017}]{Kern_2017}
Kern N.~S.,  Liu A.,  Parsons A.~R.,  Mesinger A.,   Greig B.,  2017, \mn@doi
  [The Astrophysical Journal] {10.3847/1538-4357/aa8bb4}, 848, 23

\bibitem[\protect\citeauthoryear{Kingma \& Ba}{Kingma \&
  Ba}{2017}]{kingma2017adam}
Kingma D.~P.,  Ba J.,  2017, Adam: A Method for Stochastic Optimization
  (\mn@eprint {arXiv} {1412.6980})

\bibitem[\protect\citeauthoryear{{List}, {Elahi}  \& {Lewis}}{{List}
  et~al.}{2020}]{List2020}
{List} F.,  {Elahi} P.~J.,   {Lewis} G.~F.,  2020, \mn@doi [\apj]
  {10.3847/1538-4357/abbfb2}, \href
  {https://ui.adsabs.harvard.edu/abs/2020ApJ...904..153L} {904, 153}

\bibitem[\protect\citeauthoryear{{Liu} et~al.,}{{Liu} et~al.}{2019}]{LiuWP2019}
{Liu} A.,  et~al., 2019, \baas, \href
  {https://ui.adsabs.harvard.edu/abs/2019BAAS...51c..63L} {51, 63}

\bibitem[\protect\citeauthoryear{Lovell, Frenk, Eke, Jenkins, Gao  \&
  Theuns}{Lovell et~al.}{2014}]{10.1093/mnras/stt2431}
Lovell M.~R.,  Frenk C.~S.,  Eke V.~R.,  Jenkins A.,  Gao L.,   Theuns T.,
  2014, \mn@doi [Monthly Notices of the Royal Astronomical Society]
  {10.1093/mnras/stt2431}, 439, 300

\bibitem[\protect\citeauthoryear{Mangena, Hassan  \& Santos}{Mangena
  et~al.}{2020}]{Mangena_2020}
Mangena T.,  Hassan S.,   Santos M.~G.,  2020, \mn@doi [Monthly Notices of the
  Royal Astronomical Society] {10.1093/mnras/staa750}, 494, 600–606

\bibitem[\protect\citeauthoryear{Mellema, Iliev, Pen  \& Shapiro}{Mellema
  et~al.}{2006}]{Mellema_2006}
Mellema G.,  Iliev I.~T.,  Pen U.-L.,   Shapiro P.~R.,  2006, \mn@doi [Monthly
  Notices of the Royal Astronomical Society]
  {10.1111/j.1365-2966.2006.10919.x}, 372, 679–692

\bibitem[\protect\citeauthoryear{Mesinger \& Furlanetto}{Mesinger \&
  Furlanetto}{2007}]{Mes07}
Mesinger A.,  Furlanetto S.,  2007, The Astrophysical Journal, 669, 663

\bibitem[\protect\citeauthoryear{Mesinger, Furlanetto  \& Cen}{Mesinger
  et~al.}{2011}]{Mesinger2010}
Mesinger A.,  Furlanetto S.,   Cen R.,  2011, \mn@doi [Monthly Notices of the
  Royal Astronomical Society] {10.1111/j.1365-2966.2010.17731.x}, 411, 955

\bibitem[\protect\citeauthoryear{{Mohan}, {Scaife}, {Porter}, {Walmsley}  \&
  {Bowles}}{{Mohan} et~al.}{2022}]{2022arXiv220101203M}
{Mohan} D.,  {Scaife} A. M.~M.,  {Porter} F.,  {Walmsley} M.,   {Bowles} M.,
  2022, arXiv e-prints, \href
  {https://ui.adsabs.harvard.edu/abs/2022arXiv220101203M} {p. arXiv:2201.01203}

\bibitem[\protect\citeauthoryear{Moore, Ghigna, Governato, Lake, Quinn, Stadel
  \& Tozzi}{Moore et~al.}{1999}]{Moore_1999}
Moore B.,  Ghigna S.,  Governato F.,  Lake G.,  Quinn T.,  Stadel J.,   Tozzi
  P.,  1999, \mn@doi [The Astrophysical Journal] {10.1086/312287}, 524,
  L19–L22

\bibitem[\protect\citeauthoryear{Murray, Greig, Mesinger, Muñoz, Qin, Park  \&
  Watkinson}{Murray et~al.}{2020}]{Murray2020}
Murray S.~G.,  Greig B.,  Mesinger A.,  Muñoz J.~B.,  Qin Y.,  Park J.,
  Watkinson C.~A.,  2020, \mn@doi [Journal of Open Source Software]
  {10.21105/joss.02582}, 5, 2582

\bibitem[\protect\citeauthoryear{Nair \& Hinton}{Nair \&
  Hinton}{2010}]{10.5555/3104322.3104425}
Nair V.,  Hinton G.~E.,  2010, in Proceedings of the 27th International
  Conference on International Conference on Machine Learning. ICML'10.
Omnipress, Madison, WI, USA, p. 807–814

\bibitem[\protect\citeauthoryear{Palanque-Delabrouille, Y{\`{e}}che,
  Schöneberg, Lesgourgues, Walther, Chabanier  \&
  Armengaud}{Palanque-Delabrouille et~al.}{2020}]{Palanque_Delabrouille_2020}
Palanque-Delabrouille N.,  Y{\`{e}}che C.,  Schöneberg N.,  Lesgourgues J.,
  Walther M.,  Chabanier S.,   Armengaud E.,  2020, \mn@doi [Journal of
  Cosmology and Astroparticle Physics] {10.1088/1475-7516/2020/04/038}, 2020,
  038

\bibitem[\protect\citeauthoryear{Parsons et~al.,}{Parsons
  et~al.}{2010}]{Parsons_2010}
Parsons A.~R.,  et~al., 2010, \mn@doi [The Astronomical Journal]
  {10.1088/0004-6256/139/4/1468}, 139, 1468

\bibitem[\protect\citeauthoryear{Pober et~al.,}{Pober
  et~al.}{2013}]{Pober_2013}
Pober J.~C.,  et~al., 2013, \mn@doi [The Astronomical Journal]
  {10.1088/0004-6256/145/3/65}, 145, 65

\bibitem[\protect\citeauthoryear{Pober et~al.,}{Pober
  et~al.}{2014}]{Pober_2014}
Pober J.~C.,  et~al., 2014, \mn@doi [The Astrophysical Journal]
  {10.1088/0004-637x/782/2/66}, 782, 66

\bibitem[\protect\citeauthoryear{{Prelogovi{\'c}}, {Mesinger}, {Murray},
  {Fiameni}  \& {Gillet}}{{Prelogovi{\'c}}
  et~al.}{2022}]{prelogovic2021machine}
{Prelogovi{\'c}} D.,  {Mesinger} A.,  {Murray} S.,  {Fiameni} G.,   {Gillet}
  N.,  2022, \mn@doi [\mnras] {10.1093/mnras/stab3215}, \href
  {https://ui.adsabs.harvard.edu/abs/2022MNRAS.509.3852P} {509, 3852}

\bibitem[\protect\citeauthoryear{Ramachandran, Zoph  \& Le}{Ramachandran
  et~al.}{2017}]{Swish}
Ramachandran P.,  Zoph B.,   Le Q.~V.,  2017, CoRR, abs/1710.05941

\bibitem[\protect\citeauthoryear{Reddi, Kale  \& Kumar}{Reddi
  et~al.}{2019}]{AMSGrad}
Reddi S.~J.,  Kale S.,   Kumar S.,  2019, CoRR, abs/1904.09237

\bibitem[\protect\citeauthoryear{{Sabiu}, {Kadota}, {Asorey}  \&
  {Park}}{{Sabiu} et~al.}{2021}]{2021FDM_CNN}
{Sabiu} C.~G.,  {Kadota} K.,  {Asorey} J.,   {Park} I.,  2021, arXiv e-prints,
  \href {https://ui.adsabs.harvard.edu/abs/2021arXiv210807972S} {p.
  arXiv:2108.07972}

\bibitem[\protect\citeauthoryear{Saxena, Majumdar, Kamran  \& Viel}{Saxena
  et~al.}{2020}]{Saxena_2020}
Saxena A.,  Majumdar S.,  Kamran M.,   Viel M.,  2020, \mn@doi [Monthly Notices
  of the Royal Astronomical Society] {10.1093/mnras/staa1768}, 497, 2941–2953

\bibitem[\protect\citeauthoryear{Schmit \& Pritchard}{Schmit \&
  Pritchard}{2017}]{Schmit_2017}
Schmit C.~J.,  Pritchard J.~R.,  2017, \mn@doi [Monthly Notices of the Royal
  Astronomical Society] {10.1093/mnras/stx3292}, 475, 1213–1223

\bibitem[\protect\citeauthoryear{Shimabukuro \& Semelin}{Shimabukuro \&
  Semelin}{2017}]{Shimabukuro_2017}
Shimabukuro H.,  Semelin B.,  2017, \mn@doi [Monthly Notices of the Royal
  Astronomical Society] {10.1093/mnras/stx734}, 468, 3869–3877

\bibitem[\protect\citeauthoryear{Simonyan, Vedaldi  \& Zisserman}{Simonyan
  et~al.}{2014}]{simonyan2014deep}
Simonyan K.,  Vedaldi A.,   Zisserman A.,  2014, Deep Inside Convolutional
  Networks: Visualising Image Classification Models and Saliency Maps
  (\mn@eprint {arXiv} {1312.6034})

\bibitem[\protect\citeauthoryear{Sitwell, Mesinger, Ma  \& Sigurdson}{Sitwell
  et~al.}{2014}]{Sitwell_2014}
Sitwell M.,  Mesinger A.,  Ma Y.-Z.,   Sigurdson K.,  2014, \mn@doi [Monthly
  Notices of the Royal Astronomical Society] {10.1093/mnras/stt2392}, 438,
  2664–2671

\bibitem[\protect\citeauthoryear{{The HERA Collaboration} et~al.,}{{The HERA
  Collaboration} et~al.}{2021}]{HERA2021}
{The HERA Collaboration} et~al., 2021, arXiv e-prints, \href
  {https://ui.adsabs.harvard.edu/abs/2021arXiv210807282T} {p. arXiv:2108.07282}

\bibitem[\protect\citeauthoryear{Tingay et~al.,}{Tingay
  et~al.}{2013}]{Tingay_2013}
Tingay S.~J.,  et~al., 2013, \mn@doi [Publications of the Astronomical Society
  of Australia] {10.1017/pasa.2012.007}, 30

\bibitem[\protect\citeauthoryear{{Viel}, {Lesgourgues}, {Haehnelt}, {Matarrese}
   \& {Riotto}}{{Viel} et~al.}{2005}]{2005PhRvD..71f3534V}
{Viel} M.,  {Lesgourgues} J.,  {Haehnelt} M.~G.,  {Matarrese} S.,   {Riotto}
  A.,  2005, \mn@doi [\prd] {10.1103/PhysRevD.71.063534}, \href
  {https://ui.adsabs.harvard.edu/abs/2005PhRvD..71f3534V} {71, 063534}

\bibitem[\protect\citeauthoryear{Villanueva-Domingo, Gnedin  \&
  Mena}{Villanueva-Domingo et~al.}{2018}]{Villanueva_Domingo_2018}
Villanueva-Domingo P.,  Gnedin N.~Y.,   Mena O.,  2018, \mn@doi [The
  Astrophysical Journal] {10.3847/1538-4357/aa9ff5}, 852, 139

\bibitem[\protect\citeauthoryear{{Zel'Dovich}}{{Zel'Dovich}}{1970}]{1970Zel}
{Zel'Dovich} Y.~B.,  1970, \aap, \href
  {https://ui.adsabs.harvard.edu/abs/1970A&A.....5...84Z} {500, 13}

\bibitem[\protect\citeauthoryear{Zhao, Mao, Cheng  \& Wandelt}{Zhao
  et~al.}{2021}]{zhao2021simulationbased}
Zhao X.,  Mao Y.,  Cheng C.,   Wandelt B.~D.,  2021, Simulation-Based Inference
  of Reionization Parameters From 3D Tomographic 21 cm Images (\mn@eprint
  {arXiv} {2105.03344})

\bibitem[\protect\citeauthoryear{van Haarlem et~al.,}{van Haarlem
  et~al.}{2013}]{van_Haarlem_2013}
van Haarlem M.~P.,  et~al., 2013, \mn@doi [Astronomy & Astrophysics]
  {10.1051/0004-6361/201220873}, 556, A2

\makeatother
\end{thebibliography}
%%%%%%%%%%%%%%%%%%%%%%%%%%%
%%%%%%%%%%%%%%%%%%%%%%%%%%%

%%%%%%%%%%%%%%%%%%%%%%%%%%%
%%%%%%%%%%%%%%%%%%%%%%%%%%%
\appendix
%%%%%%%%%%%%%%%%%%%%%%%%%%%
%%%%%%%%%%%%%%%%%%%%%%%%%%%

%%%%%%%%%%%%%%%%%%%%%%%%%%%
\section{2D Neural Network Architectures}
\label{sec:2DNN Architectures}
%%%%%%%%%%%%%%%%%%%%%%%%%%%

\noindent Here we display the 2D neural network architectures as presented in Section~\ref{sec:architecture}. The summary tables for our 2D CNN, ResNet and LSTM architectures are shown in Figure~\ref{tab:2DCNN_architecture}, \ref{tab:LSTM_architecture}, \ref{tab:ResNet_architecture} and , respectively. The best-performing 3D CNN architecture is shown in Table~\ref{tab:3DCNN_architecture}. 

%%%%%%%%%%%%%%%%%%%%%%%%%%%
\renewcommand{\arraystretch}{1.4}
\setlength{\tabcolsep}{8pt}
\begin{table}
    \centering
    \begin{tabular}{ll}
         Layer&Shape\\ \hline
         Input Layer &(140,2350,1)
         \\3x3 Conv2D &(138,2348,32)
         \\1x1 Zero Padding &(140,2350,32)
         \\3x3 Conv2D &(138,2348,32)
         \\2x2 Max Pooling &(69,1174,32)
         \\3x3 Conv2D &(67,1172,64)
         \\1x1 Zero Padding &(69,1174,64)
         \\3x3 Conv2D &(67,1172,64)
         \\2x2 Max Pooling &(33,586,64)
         \\3x3 Conv2D &(31,584,128)
         \\1x1 Zero Padding &(33,586,128)
         \\3x3 Conv2D &(31,584,128)
         \\Global Average Pooling &(128)
         \\Dense &(128)
         \\Dense &(128)
         \\Dense &(128)
         \\Dense &(82)
         \\Dense &(6)\\ \hline
         Number of Parameters: 347,044
    \end{tabular}
    \caption{2D CNN Model Summary.}
    \label{tab:2DCNN_architecture}
\end{table}
%%%%%%%%%%%%%%%%%%%%%%%%%%%

%%%%%%%%%%%%%%%%%%%%%%%%%%%
\begin{table}
    \centering
    \begin{tabular}{ll}
         Layer&Shape\\ \hline
         Input Layer &(39,140,140,1)
         \\Time Distributed 5x5 Conv2D &(39,136,136,64)
         \\Time Distributed 2x2 Max Pooling &(39,68,68,64)
         \\Time Distributed 3x3 Conv2D &(39,66,66,128)
         \\Time Distributed Global Average Pooling &(39,128)
         \\Spatial Dropout 0.1 &(39,128)
         \\LSTM &(300)
         \\Dense &(300)
         \\Dense &(120)
         \\Dense &(120)
         \\Dense &(82)
         \\Dense &(6)\\ \hline
         Number of Parameters: 741,680
    \end{tabular}
    \caption{LSTM Model Summary. The time-distributed layers of the LSTM apply the same convolutional layers to each time step.}
    \label{tab:LSTM_architecture}
\end{table}
%%%%%%%%%%%%%%%%%%%%%%%%%%%

%%%%%%%%%%%%%%%%%%%%%%%%%%%
\renewcommand{\arraystretch}{1.4}
\setlength{\tabcolsep}{8pt}
\begin{table}
    \centering
    \begin{tabular}{ll}
         Layer&Shape\\ \hline
         Input Layer &(140,2350,1)
         \\1. 6x6 Conv2D &(135,2345,32)
         \\2. 3x3 Conv2D &(135,2345,32)
         \\3. 3x3 Conv2D &(135,2345,32)
         \\4. Add 1+3 &(135,2345,32)
         \\5. 3x3 Conv2D &(135,2345,32)
         \\6. 3x3 Conv2D &(135,2345,32)
         \\7. Add 4+6 &(135,2345,32)
         \\8. Batch Normalisation &(135,2345,32)
         \\9. 3x3 Conv2D &(68,1173,64)
         \\10. 3x3 Conv2D &(68,1173,64)
         \\11. Add 8+10 &(68,1173,64)
         \\12. 3x3 Conv2D &(68,1173,64)
         \\13. 3x3 Conv2D &(68,1173,64)
         \\14. Add 11+13 &(68,1173,64)
         \\15. Batch Normalisation &(68,1173,64)
         \\16. 3x3 Conv2D &(34,587,128)
         \\17. 3x3 Conv2D &(34,587,128)
         \\18. Add 15+17 &(34,587,128)
         \\19. 3x3 Conv2D &(34,587,128)
         \\20. 3x3 Conv2D &(34,587,128)
         \\21. Add 18+20 &(34,587,128)
         \\22. Global Average Pooling &(128)
         \\23. Dense &(128)
         \\23. Dropout 0.2 &(128)
         \\24. Dense &(128)
         \\25. Dense &(128)
         \\25. Dense &(82)
         \\26. Dense &(6)\\ \hline
         Number of Parameters: 755,492
    \end{tabular}
    \caption{Residual Network (ResNet) Model Summary. Convolutional layers apart from the first one include zero padding to keep their size to the original one. Layers 9 and 16 use stride of 2 for down-sampling. A convolutional layer with kernel size 1 and stride 2 is applied to layers 8 and 15 before adding them to the down-sampled layers in layer 11 and 18.}
    \label{tab:ResNet_architecture}
\end{table}
%%%%%%%%%%%%%%%%%%%%%%%%%%%

%%%%%%%%%%%%%%%%%%%%%%%%%%%
\section{Transfer Learning Scatter plots for astrophysical parameters}
\label{sec:TransferAppendix}
%%%%%%%%%%%%%%%%%%%%%%%%%%%

\begin{figure*}
\includegraphics[width=0.329\textwidth]{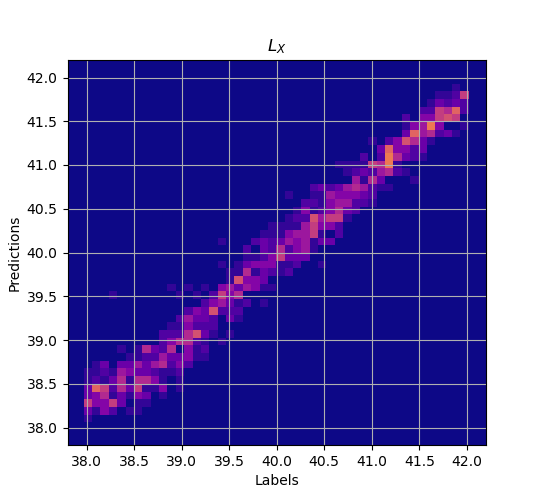}
\includegraphics[width=0.405\textwidth]{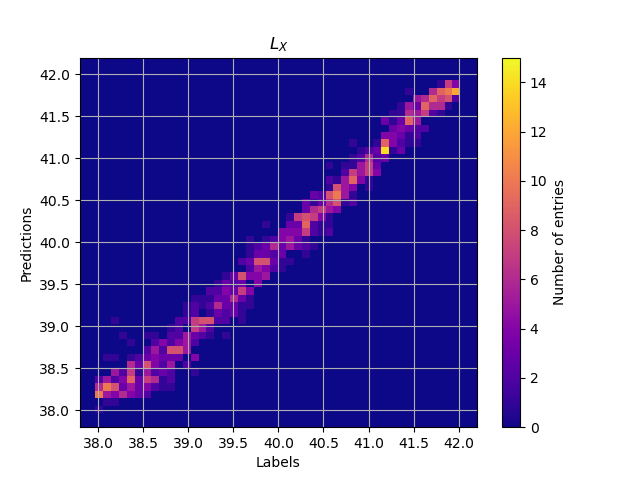}
\includegraphics[width=0.329\textwidth]{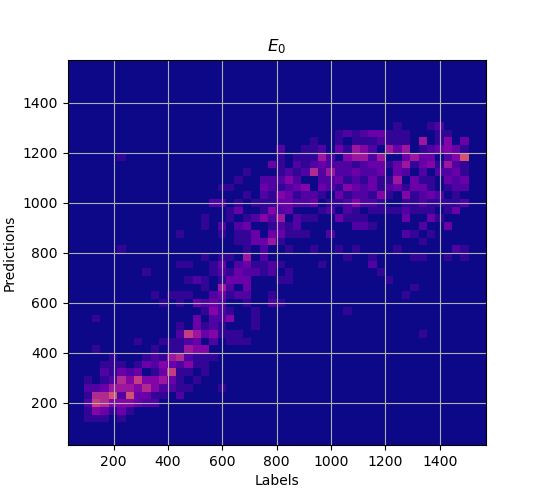}
\includegraphics[width=0.405\textwidth]{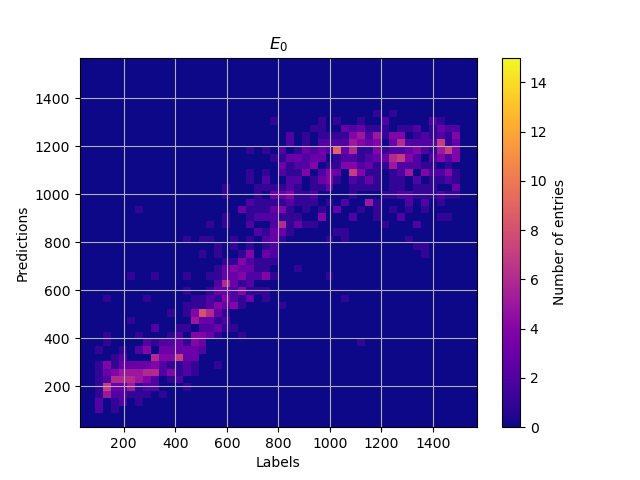}
\includegraphics[width=0.329\textwidth]{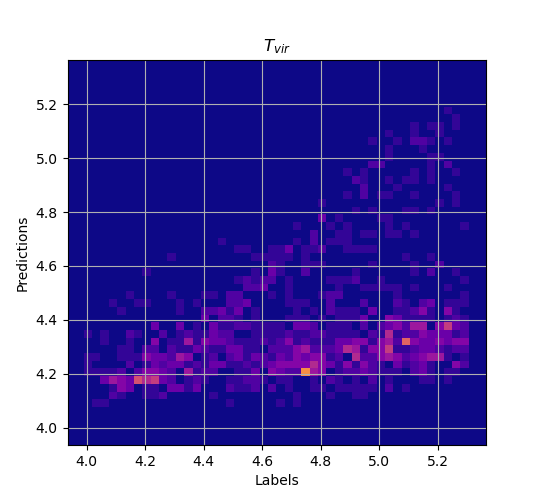}
\includegraphics[width=0.405\textwidth]{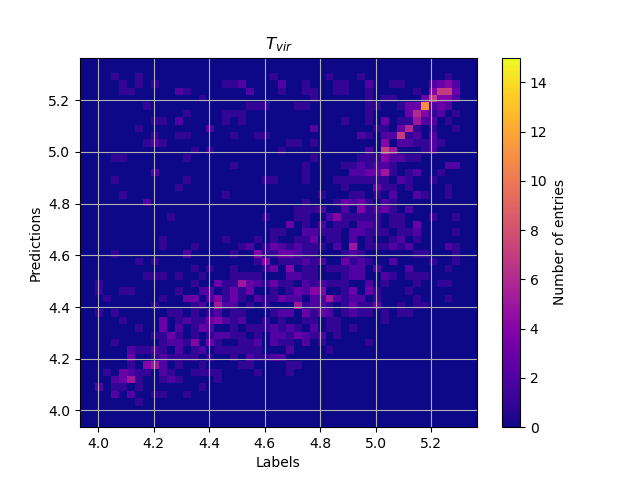}
\includegraphics[width=0.329\textwidth]{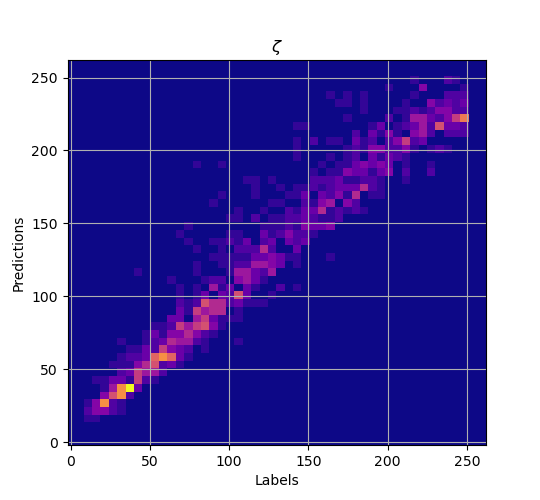}
\includegraphics[width=0.405\textwidth]{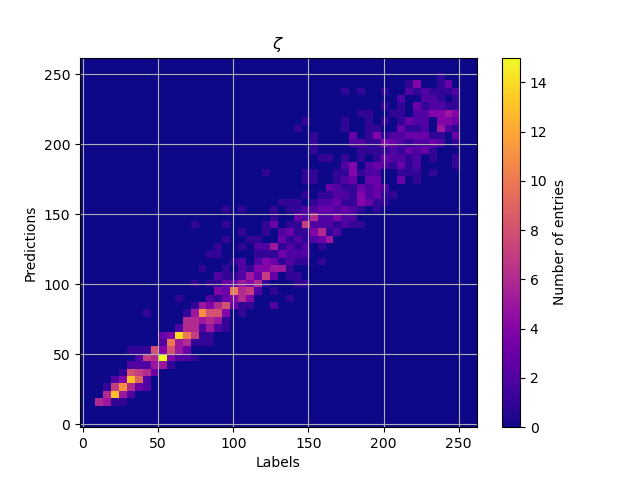}
\caption{Scatter plots for transfer learning astrophysical parameters according to Section~\ref{sec:TransferLearning}. Left panels: 3D CNN predictions on simulations after being trained on opt mock light-cones; right panels: 3D CNN predictions on opt mock light-cones after being trained on simulations.}
\label{fig:TransferLearningAppendix}
\end{figure*}

\noindent Here we briefly discuss the scatter for the astrophysical parameters obtained for transfer learning between bare simulated light-cones and mock observational ones; see Figure~\ref{fig:TransferLearningAppendix} for the individual scatter plots. For methodology we refer to Section~\ref{sec:TransferLearning}. The scatter plots for $L_\mathrm{X}$, $E_0$ and $\zeta$ are barely affected by transfer learning from an opt mock trained 3D CNN to bare simulations (left) and vice versa (right). However in the former case we get a slightly broader scatter for all parameters. Regarding $T_\mathrm{vir}$ the majority of predictions is highly scattered. We again note that the parameters which strongly influence late heating and reionisation, where the difference between opt mock light-cones and bare simulations is small, yield good results in transfer learning. For $T_\mathrm{vir}$, which directly affects star formation up to early heating, the 3D CNN fails to transfer from simulations to mock light-cones, and vice versa.

%%%%%%%%%%%%%%%%%%%%%%%%%%%
%%%%%%%%%%%%%%%%%%%%%%%%%%%

%%%%%%%%%%%%%%%%%%%%%%%%%%%
\section{Network Interpretation}
\label{sec:InterpretationAppendix}
%%%%%%%%%%%%%%%%%%%%%%%%%%%

%%%%%%%%%%%%%%%%%%%%%%%%%%%
\begin{figure*}
\includegraphics[width=0.95\textwidth]{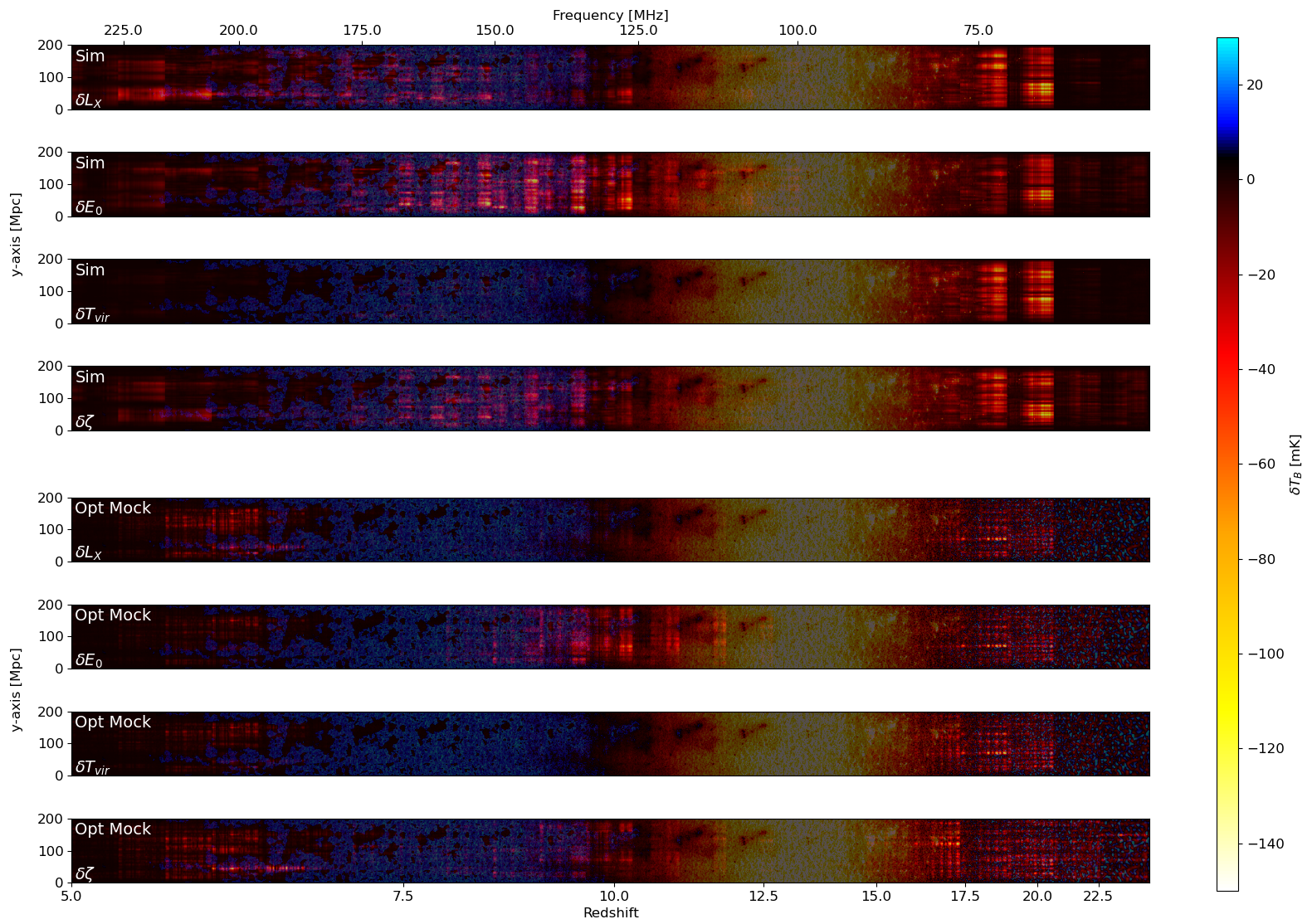}
\caption{Saliency maps for our best-performing 3D CNN after being trained on simulations (top four slices) and opt mock light-cones (bottom four slices), respectively, in both cases for the remaining set of astrophysical parameters, X-ray luminosity $L_\mathrm{X}$, energy threshold $E_0$, virial temperature $T_\mathrm{vir}$ and ionising efficiency $\zeta$ (from top to bottom). For a more detailed description of how these maps were created see Section~\ref{sec:saliency}, and for the saliency maps of WDM mass and the DM density parameter see Figure~\ref{fig:Saliency}.}
\label{fig:SaliencyAppendix}
\end{figure*}
%%%%%%%%%%%%%%%%%%%%%%%%%%%

\noindent We here show in Figure~\ref{fig:SaliencyAppendix} saliency maps for our set of astrophysical parameters. For details on the creation of saliency maps as well as the saliency maps for $m_\mathrm{WDM}$ and $\Omega_\mathrm{m}$ see Section~\ref{sec:saliency}. As can be seen in Figure~\ref{fig:SaliencyAppendix}, a behaviour similar as for the DM parameters can be observed. For mock light-cones (bottom four panels) the 3D CNN focuses on the transition areas, i.e. the onset of heating, the transition between heating and reionisation, and the late states of reionisation, and thus is better physically motivated focusing on when the largest changes in the IGM structure occur. For example for $E_0$ strong focus lies on the transition between heating and reionisation, as expected for an X-ray heating parameter. Conversely, for $L_\mathrm{X}$ the same area has less influence on predictions, while the onset of heating is more important. For most parameters the neural networks focus on the onset of heating, indicating the timing of heating and the appearance of the first sources at high redshifts is an important discriminator for the networks. In comparison, for bare simulated light-cones (top four panels) broad ranges across light-cones are of importance for the astrophysical parameters.

\begin{figure*}
\includegraphics[width=0.3\textwidth]{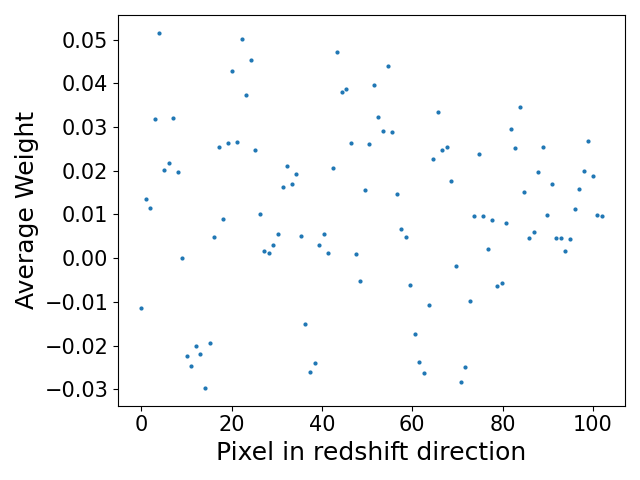}
\includegraphics[width=0.3\textwidth]{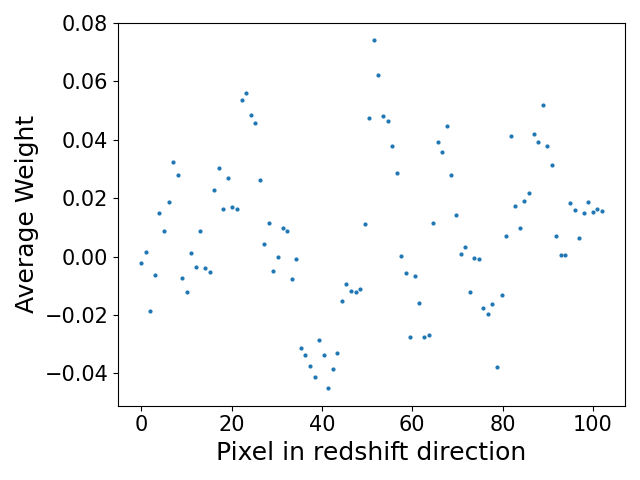}
\includegraphics[width=0.3\textwidth]{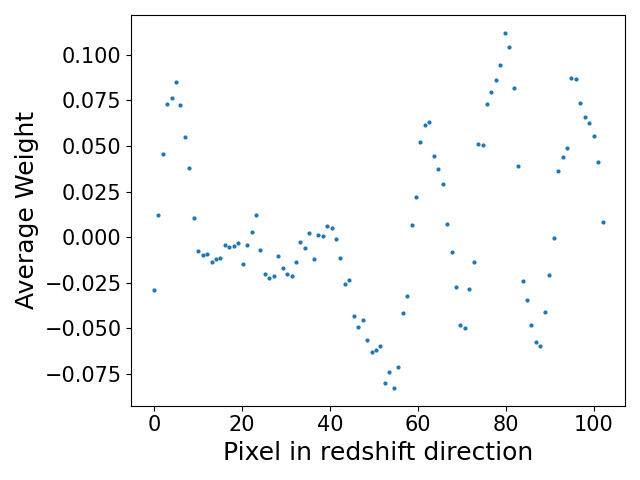}
\includegraphics[width=0.3\textwidth]{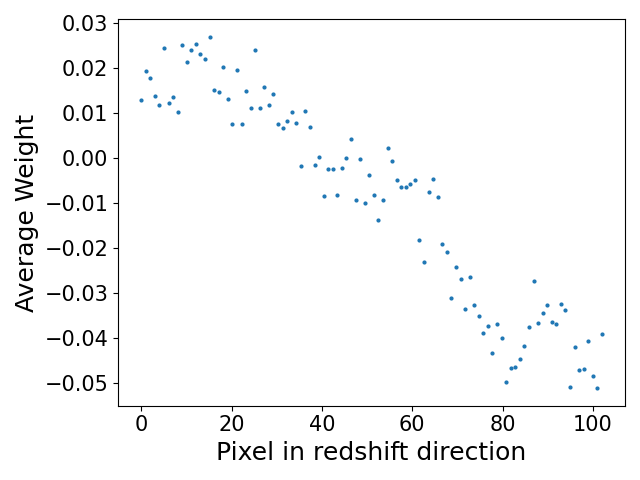}
\includegraphics[width=0.3\textwidth]{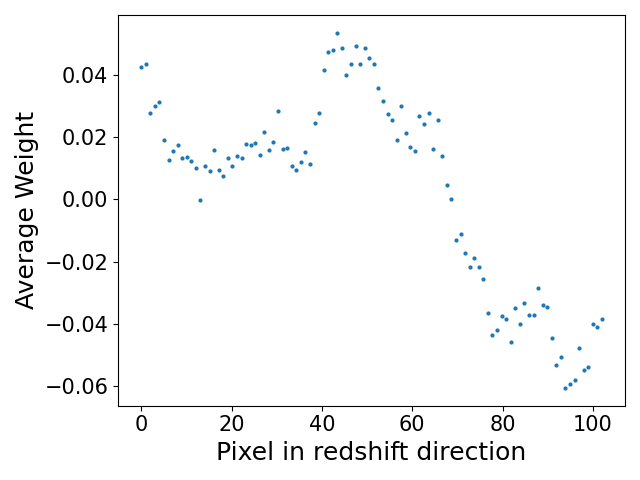}
\includegraphics[width=0.3\textwidth]{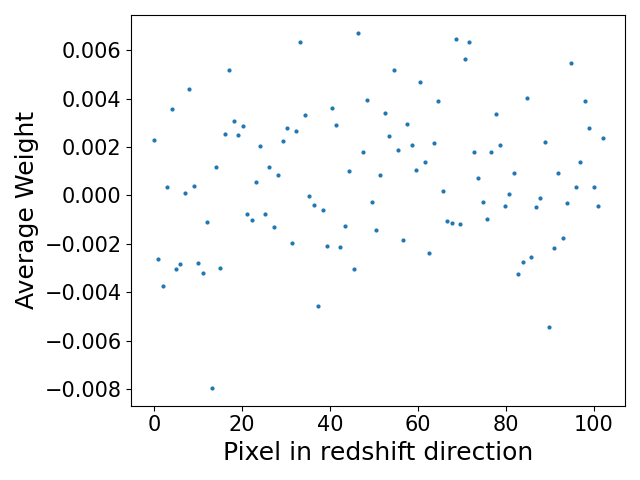}
\caption{Six representative examples for filters from the first layer of our 3D CNN network after training on opt mock light-cones; the weights are averaged over the two spatial dimensions of each filter to be plotted against the direction in redshift.}

\label{fig:Filter}
\end{figure*}

In addition to saliency maps, for interpretation purposes we show a representative set of six filters for our 3D CNN trained on opt mock light-cones in Figure~\ref{fig:Filter}; see also Section~\ref{sec:Filter}. Three main types are present in the filters, mountains and troughs, rises and falls that seem to map general trends in brightness temperature, and more complicated structures. Typical representations of mountain and trough structures are shown in the top three panels. A common feature they share is the average weight of the filters being close to zero, potentially useful to identify small scale fluctuations. The most prominent fluctuations can be found during early heating, late reionisation and the transition between heating and reionisation, which are also the main focus of the saliency maps. A combination of troughs and mountains, tracing the shape of areas of emission or absorption together with rising or falling structures (bottom left) allows a network to both model fluctuations and large-scale transitions. More complicated structures (bottom right), but with an average weight again close to zero, is less useful in determining absolute values, but might rather identify substructure. 
%%%%%%%%%%%%%%%%%%%%%%%%%%%
%%%%%%%%%%%%%%%%%%%%%%%%%%%

\bsp
\end{document}